\documentclass[aps,pre, superscriptaddress, reprint, amsmath, amssymb]{revtex4-2}

\pdfoutput=1
\usepackage[utf8]{inputenc}
\usepackage{bm}
\usepackage[english]{babel}
\usepackage[T1]{fontenc}
\usepackage{mathrsfs}
\usepackage[retainorgcmds]{IEEEtrantools}
\usepackage{amsmath}
\usepackage{amssymb}
\usepackage{color}
\usepackage{amsfonts}
\usepackage{times,txfonts}
\usepackage{nicefrac}
\usepackage[colorlinks=true,linkcolor=blue,urlcolor=blue,citecolor=blue,pdfusetitle]{hyperref}

\renewcommand{\r}{\hat{\rho}}
\newcommand{\dd}{\mathrm{d}}
\newcommand{\ii}{\ensuremath{\mathrm{i}}}
\newcommand{\ee}{\ensuremath{\mathrm{e}}}

\newcommand{\pd}[2]{\ensuremath{\frac{\partial #1}{\partial #2}}}

\usepackage{float}
\usepackage{epsfig} 
\usepackage{subfigure}
\graphicspath{{Images/}}

\usepackage{physics}
\usepackage{braket}
\usepackage{amssymb}
\usepackage[amssymb]{SIunits}

\usepackage{xcolor}

\newcommand{\Conv}{%
  \mathop{\scalebox{1.5}{\raisebox{-0.2ex}{$\circledast$}}
  }
}

\begin{document}

\title{Powering an autonomous clock with quantum electromechanics}

\author{Oisín Culhane}
\email{oculhane@tcd.ie}
\affiliation{School of Physics, Trinity College Dublin, Dublin 2, Ireland}
\author{Michael J. Kewming}
\email{kewmingm@tcd.ie}
\affiliation{School of Physics, Trinity College Dublin, Dublin 2, Ireland}
\author{Alessandro Silva}
\email{alessandro.silva@sissa.it}
\affiliation{SISSA, Via Bonomea 265, I-34135 Trieste, Italy}
\author{John Goold}
\email{gooldj@tcd.ie}
\affiliation{School of Physics, Trinity College Dublin, Dublin 2, Ireland}
\affiliation{Trinity Quantum Alliance, Unit 16, Trinity Technology and Enterprise Centre, Pearse Street, Dublin 2, D02YN67}
\author{Mark T. Mitchison}
\email{mark.mitchison@tcd.ie}
\affiliation{School of Physics, Trinity College Dublin, Dublin 2, Ireland}
\affiliation{Trinity Quantum Alliance, Unit 16, Trinity Technology and Enterprise Centre, Pearse Street, Dublin 2, D02YN67}

\begin{abstract}
We theoretically analyse an autonomous clock comprising a nanoelectromechanical system, which undergoes self-oscillations driven by electron tunnelling. The periodic mechanical motion behaves as the clockwork, similar to the swinging of a pendulum, while induced oscillations in the electrical current can be used to read out the ticks. We simulate the dynamics of the system in the quasi-adiabatic limit of slow mechanical motion, allowing us to infer statistical properties of the clock's ticks from the current auto-correlation function. The distribution of individual ticks exhibits a tradeoff between accuracy, resolution, and dissipation, as expected from previous literature. Going beyond the distribution of individual ticks, we investigate how clock accuracy varies over different integration times by computing the Allan variance. We observe non-monotonic features in the Allan variance as a function of time and applied voltage, which can be explained by the presence of temporal correlations between ticks. These correlations are shown to yield a precision advantage for timekeeping over the timescales that the correlations persist. Our results illustrate the non-trivial features of the tick series produced by nanoscale clocks, and pave the way for experimental investigation of clock thermodynamics using nanoelectromechanical systems. 
\end{abstract}

\maketitle

\section{\label{sect:intro}Introduction}


The measurement of time is perhaps the most primitive and also the most important of all physical measurements. Not only are clocks ubiquitous in our daily lives, but they also underpin satellite navigation, magnetometry, and gravimetry~\cite{stray_quantum_2022, bothwell_resolving_2022}, among many other applications in metrology. However, despite the spectacular precision that is now routinely achieved with atomic clocks, fundamental gaps in our understanding of time at the quantum level still remain.

One fruitful and long-standing~\cite{salecker_quantum_1958,peres_measurement_1980,page_evolution_1983} strategy to address this problem is to formulate and study models of clocks as dynamical systems. Recent work in this direction has shown that basic physical considerations, stemming from information theory on the one hand~\cite{janzing_quasi-order_2003,rankovic_quantum_2015,woods_autonomous_2019,woods_autonomous_2021,woods_quantum_2022} and thermodynamics on the other~\cite{Erker_2017,Barato_2016, Milburn2020,Schwarzhans2021}, strongly constrain both the internal structure and the achievable performance of timekeeping devices. For example, it is now understood that accurate clocks must produce a lot of entropy~\cite{Erker_2017,Milburn2020}, quantum clocks can be more accurate than classical clocks with the same number of accessible states~\cite{woods_quantum_2022,woods_autonomous_2021,dost_quantum_2023}, and that there exists a tradeoff between a clock's accuracy and its timing resolution~\cite{meier_fundamental_2023}. Similar constraints on temporal precision known as thermodynamic or kinetic uncertainty relations have been discovered in other, seemingly unrelated contexts, such as biomolecular systems~\cite{barato_thermodynamic_2015,gingrich_dissipation_2016, garrahan_simple_2017,terlizzi_kinetic_2018} or mesoscopic transport setups~\cite{ptaszynski_coherence-enhanced_2018,agarwalla_assessing_2018,guarnieri_thermodynamics_2019}. Beyond their foundational importance, moreover, the physical limits of timekeeping have consequences for the fidelity of time-dependent control protocols arising in quantum thermodynamics~\cite{malabarba_clock-driven_2015, frenzel_quasi-autonomous_2016,woods_autonomous_2023} and quantum computation~\cite{ball_role_2016,xuereb_impact_2023}, among other applications.

Here we focus on ticking clocks, which produce a continuous time reference to an external observer, as opposed to stopwatch-like systems (e.g.~the Larmor clock~\cite{hauge_tunneling_1989}) that measure only a single time interval. Any quantum ticking clock comprises two distinct parts: the clockwork and the register~\cite{rankovic_quantum_2015}. The clockwork is a non-equilibrium system whose dynamics generates the ticks, which are then recorded by the register. Separation of clockwork and register ensures that the tick readout does not disrupt the clockwork~\cite{silva_ticking_2023}. Although almost any out-of-equilibrium system can act as a clockwork, accurate timekeeping requires temporal probability concentration~\cite{Schwarzhans2021}, whereby the irreversible process that generates each tick is permitted to occur only at well-spaced time intervals. For example, while a simple hourglass exploits freefalling mass (e.g.~sand) to indicate the passage of time, the escapement of a pendulum clock allows a mass to fall incrementally --- thereby generating the next tick --- only after one full period of the swinging pendulum has elapsed~\cite{Milburn2020,pietzonka_classical_2022}. However, much of this new understanding of quantum ticking clocks has been garnered via theoretical analysis of simple toy models. The connection between the abstract theory of autonomous quantum clocks and concrete experimental implementations remains lacking.


Here, we fill this gap, by proposing and analysing an autonomous quantum clock based on nanoelectromechanical self-oscillations, which have recently been observed in the laboratory~\cite{Wen2020,Urgell2020}. A self-oscillator~\cite{jenkins_self-oscillation_2013} is a system that oscillates in the absence of periodic forcing, and therefore provides the ideal setting to realise an autonomous clockwork. In nanoelectromechanical self-oscillators, the periodic motion of a mechanical degree of freedom is driven by electron tunnelling~\cite{Bachtold2022}. This can occur either by electron shuttling~\cite{Gorelik1998,Park_2000,erbe_nanomechanical_2001,Pistolesi2006,Novotn2003,Novotn2004}, where single-electron tunnelling resonantly drives the oscillator, or via a subtler electromechanical instability that emerges from the time-delayed effect of many non-resonant tunnelling events~\cite{Blanter2004,Blanter2004err,Clerk2005,Bennett2006,Usmani2007,Wen2020,Urgell2020}. Thermodynamic aspects of both kinds of self-oscillations have recently been investigated~\cite{Wachtler_2019,Strasberg2021,Culhane_2022}, but their potential for studying the thermodynamics of timekeeping has not yet been studied in detail.

In this work, we consider an autonomous clock powered by electromechanical self-oscillations of the latter kind, motivated by the experiments reported in Refs.~\cite{Wen2020,Urgell2020}. In our model, periodic mechanical motion provides the clockwork while the ticks are registered by monitoring the induced oscillations in the electronic current. As in previous studies, the autonomous nature of the clockwork dynamics allows for a self-contained analysis of the thermodynamic resources needed to measure time~\cite{Erker_2017}. However, our work goes beyond existing literature on clock thermodynamics in at least two important respects. 

First, our electromechanical clockwork is based on the oscillations of a continuous degree of freedom, reminiscent of a pendulum. Yet, unlike a pendulum clock, or indeed any other clock considered in the literature to our knowledge, these oscillations are driven by measurement backaction associated with the tick readout itself. By contrast, existing theoretical models of quantum clocks consider the tick generation and readout to arise from distinct irreversible processes~\cite{Erker_2017,Schwarzhans2021,woods_quantum_2022,woods_autonomous_2021,dost_quantum_2023,meier_fundamental_2023,silva_ticking_2023}, while experimental work on the thermodynamics of timekeeping has either lain in a purely classical regime where measurement backaction is negligible~\cite{Pearson2021} or focussed purely on the role of measurement backaction in a non-autonomous setting~\cite{he_quantum_2023}. Our setup is also distinct from Ref.~\cite{manikandan_autonomous_2023}, which considered a quantum clock powered by measurements that are unrelated to tick readout. 

Second, we go beyond the performance measures introduced in previous works~\cite{Erker_2017,woods_quantum_2022} by evaluating the Allan variance~\cite{Allan_1966}, which quantifies clock accuracy over different integration times. At intermediate timescales, our results corroborate the general expectation that higher accuracy entails more dissipation. However, at longer timescales this trend reverses. To explain this crossover in the Allan variance, we show that the non-trivial dynamics of the clockwork generates persistent temporal correlations between the clock's ticks. Our results thus call for further research to understand both the origin and consequences of such temporal correlations, which have been largely neglected in previous studies of autonomous quantum clocks.

\begin{figure}
	\begin{center}
		\includegraphics[width=\columnwidth]{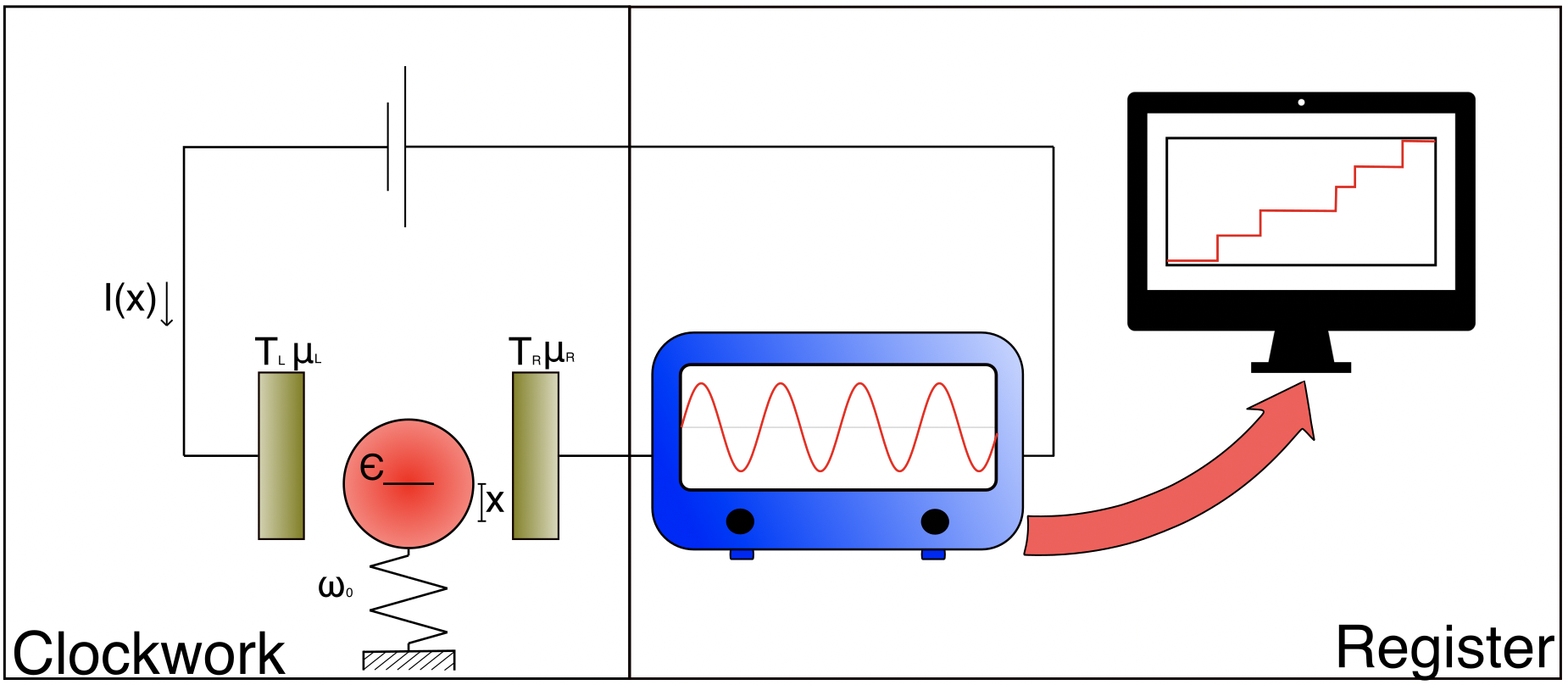}
		\caption{Schematic of the setup.  A central system, here a quantum dot, is coupled to two electrodes each at a temperature $T$, with the left electrode having a chemical potential of $\mu_L$ and the right a chemical potential of $\mu_R$. The quantum dot is coupled to simple harmonic oscillator mode. Under an appropriate static voltage bias, the current flowing through the system drives self-sustained mechanical oscillations, which in turn induces an oscillatory component in the current. A series of ticks are recorded by counting oscillations in the output current. \label{fig:Schematic}}
	\end{center}
\end{figure}

An outline of the manuscript is as follows. We first introduce our model for an electromechanical clock (Sec.~\ref{sec:setup}), and explain how we model the clockwork's nonequilibrium dynamics (Sec.~\ref{sec:langevin}) and the readout of ticks from oscillations induced in the electrical current (Sec.~\ref{sec:current_model}). Our modelling is based on the quasi-adiabatic assumption, i.e.~that the mechanical motion is slow compared to electronic relaxation, which is well satisfied in current experiments~\cite{Wen2020,Urgell2020}. Then we examine the induced current oscillations in detail, identifying an interesting separation of timescales between amplitude and phase fluctuations (Sec.~\ref{sec:corr_func}). Next, we analyse the distribution of individual ticks of the clock (Sec.~\ref{sec:wtd}) and find a tradeoff between accuracy, resolution, and entropy production, as expected from previous work~\cite{Erker_2017}. Finally, we show that the Allan variance exhibits a non-trivial dependence on integration time and applied voltage (Sec.~\ref{sect:allan_var}), which can be understood in terms of correlations between the clock's ticks that we quantify explicitly  (Sec.~\ref{sec:tick_correlations}). We conclude with a summary and a brief outlook on future work (Sec.~\ref{sect:disc}). The reduced Planck constant, the Boltzmann constant, and the electron charge are all set to unity throughout this work. 

\section{\label{sect:model}Model}

\subsection{Description of the setup}
\label{sec:setup}

We consider an electromechanical system similar to the model outlined in \cite{Culhane_2022}. A schematic of the setup is shown in Fig.~\ref{fig:Schematic}. The clockwork consists of a resonant-level model coupled to a harmonic oscillator. The oscillator motion is in the direction perpendicular to the current flow (parallel to the plane defined by the leads), as shown in Fig.~\ref{fig:Schematic}. The Hamiltonian of the clockwork is given by a sum of the following terms:
\begin{align}
\label{ham_quantum_dot}
    & \hat{H}_S = \epsilon \hat{c}^\dagger \hat{c}, \\
\label{ham_baths}
    & \hat{H}_B = \sum_k \left(\Omega_{kL}\hat{d}^\dagger_{kL}\hat{d}_{kL} + \Omega_{kR}\hat{d}^\dagger_{kR}\hat{d}_{kR} \right) ,\\
\label{ham_tunnelling}
    & \hat{H}_T = \sum_k \left( g_{kL}\left[\hat{c}^\dagger \hat{d}_{kL} + \hat{d}^\dagger_{kL} \hat{c}\right] + g_{kR}\left[\hat{c}^\dagger \hat{d}_{kR} + \hat{d}^\dagger_{kR} \hat{c}\right] \right),\\
\label{ham_vibrational}
   & \hat{H}_V = \frac{\hat{p}^2}{2m} + \frac{m\omega_0^2 \hat{x}^2}{2} - F\hat{n}\hat{x}.
\end{align}
Here, $\hat{H}_S$ is the Hamiltonian of a single-level fermionic quantum dot, described by its energy $\epsilon$ and the annihilation operator $\hat{c}$. The leads coupled to the quantum dot are defined by the Hamiltonian $\hat{H}_B$. Each lead is described by an infinite sum of noninteracting Fermi modes with energy $\Omega_{k\alpha}$ and annihilation operator $\hat{d}_{k\alpha}$, where $\alpha = L,R$ denotes the left or right lead. The tunnelling of electrons between the dot and the leads is described by $\hat{H}_T$, with $g_{k\alpha}$ the coupling strength. The vibrational Hamiltonian $\hat{H}_V$ describes a single-mode mechanical oscillator constrained to move in one dimension, with momentum operator $\hat{p}$, position operator $\hat{x}$, natural oscillation frequency $\omega_0$, and  mass $m$. The final term in $\hat{H}_V$ describes the electromechanical interaction between the quantum dot and the harmonic oscillator, where $F$ is the force per unit charge and $\hat{n}$ is the excess charge present on the quantum dot, i.e.~$\hat{n} = \hat{c}^\dagger\hat{c}-N_0$, with $N_0 = \langle\hat{c}^\dagger\hat{c}\rangle_{F=0}$ being the average charge on the oscillator in the absence of the electromechanical coupling. This kind of coupling can be induced, for example, by means of a nearby gate electrode held at a constant potential, which produces an electrostatic force on the oscillator depending on the number of charges present in the device~\cite{Wen2020} (e.g.~in Fig.~\ref{fig:Schematic} this force would act vertically).

The leads are initially at equilibrium with equal inverse temperature $\beta$ but different chemical potentials, $\mu_L\neq \mu_R$, with $V=\mu_L-\mu_R$ denoting the voltage bias. The  tunnelling rate for each lead is determined by its spectral density
\begin{equation}
\label{spectral_density}
\kappa_\alpha(E)  = 2\pi \sum_k g_{k\alpha}^2 \delta(E - \Omega_{k\alpha}).
\end{equation}
Self-sustained oscillations require an energy-dependent spectral density~\cite{Clerk2005}. As in our previous work~\cite{Culhane_2022}, we assume the simple Lorentzian form
\begin{equation}
	\label{tunnel_rate}
    \kappa_\alpha(E) = \frac{\Gamma\delta^2}{(E-\omega_\alpha)^2+\delta^2},
\end{equation}
which models a single band of states with bandwidth $\delta$ centred on energy $\omega_\alpha$, with
$\Gamma$ representing the overall tunnelling rate. We focus on the quasi-adiabatic limit, where the relaxation rate of the electronic degrees of freedom are on a much faster timescale than the characteristic timescale of the oscillator. This is tantamount to the conditions
\begin{equation}
\label{adiabatic_condition}
    \min(\beta^{-1},V), \Gamma \gg \omega_0,\lambda,
\end{equation}
where $\lambda = F/\sqrt{m\omega_0}$ is the effective electromechanical coupling strength. These conditions are well satisfied in typical experiments~\cite{Wen2020,Urgell2020} and also by our specific parameter choices below. Note that our minimal model neglects charging effects due to Coulomb interaction as well as intrinsic mechanical damping of the oscillator, since neither of these effects are crucial for the functioning of the clock. 

\subsection{Langevin equation for the clockwork dynamics}
\label{sec:langevin}

When the quasi-adiabatic conditions~\eqref{adiabatic_condition} hold, the electronic degrees of freedom can be adiabatically eliminated to obtain a description for the oscillator alone. Full details can be found in the Supplemental Material of Ref.~\cite{Culhane_2022}, but we briefly sketch the derivation here because it is serves as a precursor for the modelling of tick readout described in Sec.~\ref{sec:current_model}. 


We divide the Hamiltonian into two parts as $\hat{H}=\hat{H}_{\text{el}}+\hat{H}_V$, where $\hat{H}_{\text{el}}=\hat{H}_S+\hat{H}_B+\hat{H}_T$ describes the electronic degrees of freedom and $\hat{H}_V$ describes the oscillator dynamics including the electromechanical coupling. We now introduce the Wigner transform of the density matrix
\begin{equation}
	\label{Wigner_transform}
    \hat{\rho}(x,p)=\frac{1}{\pi}\int\text{d}y\bra{x+y}\hat{\rho}\ket{x+y}\text{e}^{-i2yp},
\end{equation}
where $\ket{x}$ is a position eigenstate for the oscillator, so that $\hat{\rho}(x,p)$ acts as a quantum operator on the electronic Hilbert space. The Wigner function of the oscillator follows from $W(x,p)=\text{tr}[\hat{\rho}(x,p)]$, while the electronic state is given by $\hat{\rho}_\text{el}=\int\text{d}x\int\text{d}p\hat{\rho}(x,p)$. Using the operator correspondences for the Wigner function~\cite{GardinerZoller}, one finds that the quantum Liouville equation $\text{d}\hat{\rho}/\text{d}t=-i[\hat{H},\hat{\rho}]$ is equivalent to
\begin{equation}
	\label{Liouville_phase_space}
	 \frac{\partial}{\partial t}\hat{\rho}(x,p) = \left ( \mathcal{L}_x + \mathcal{H} + \mathcal{F}\right ) \hat{\rho}(x,p),
\end{equation}
where we define the following superoperators:
\begin{align}
	\label{L_super}
	\mathcal{L}_x\bullet &= -i\left[ \hat{H}_\text{el}-Fx\hat{n},\bullet\right],\\
		\label{H_super}
	\mathcal{H}\bullet &= \left(m\omega_0^2x-F\langle\hat{n}\rangle_x\right)\frac{\partial}{\partial p}\bullet-\frac{p}{m}\frac{\partial}{\partial x}\bullet,\\
	\label{F_super}
	\mathcal{F}\bullet&=-\frac{F}{2}\frac{\partial}{\partial p}\left\{\hat{n}-\langle\hat{n}\rangle_x,\bullet\right\}.
\end{align}
Here, $\langle\bullet\rangle_x=\text{tr}\left[\bullet\hat{\rho}_x\right]$ denotes an average with respect to the electronic stationary state, $\hat{\rho}_x$, which depends on $x$ but not on $p$, and satisfies $\mathcal{L}_x\hat{\rho}_x=0$ and $\text{tr}\left[\hat{\rho}_x\right]=1$. Physically, $\mathcal{L}_x$ represents the evolution of the electronic degrees of freedom conditioned on a particular oscillator position $x$,  $\mathcal{H}$ represents the free mechanical evolution including the mean-field electromechanical force, and $\mathcal{F}$ represents the fluctuating force on the oscillator due to deviations of the electronic charge from its steady-state mean value.

%
%

By assumption (i.e.~because of the conditions~\eqref{adiabatic_condition}), we have $\mathcal{L}_x\gg \mathcal{H},\mathcal{F}$ so that the electronic degrees of freedom are only weakly perturbed from their instantaneous steady state $\hat{\rho}_x$. This is formalised by defining, for any operator-valued function $\hat{A}(x,p)$ of the oscillator's phase-space coordinates, the projector 
\begin{equation}
	\label{projection}
    \mathcal{P}\hat{A}(x,p) = \text{tr}\left[\hat{A}(x,p)\right]\hat{\rho}_x,
\end{equation}
whose orthogonal complement is denoted by $\mathcal{Q}=1-\mathcal{P}$. The ``relevant'' part of the density matrix $\mathcal{P}\hat{\rho}(x,p)$ determines the Wigner function, since $W(x,p) = \tr \mathcal{P}\hat{\rho}(x,p)$. Thus, following the standard procedure~\cite{Gardiner}, we can perturbatively eliminate the irrelevant part, $\mathcal{Q}\hat{\rho}(x,p)$, taking into account contributions up to second order in $\mathcal{F}$ and $\mathcal{H}$ within a Born-Markov approximation (see Ref.~\cite{Culhane_2022} for details). This results in a Fokker-Planck equation for the Wigner function,  which is equivalent to the classical Langevin equation~\cite{Clerk2005,Bennett2006,Usmani2007}
\begin{equation}
\label{langevin_equation}
    m\ddot{x} + m\gamma_x\dot{x} + m\omega_0^2x = F\langle \hat{n}\rangle_x + \sqrt{D_x}\xi(t),
\end{equation}
where $\xi(t)$ is a white-noise force with zero mean, $\mathbb{E}[\xi(t)]=0$, and autocorrelation function $\mathbb{E}[\xi(t)\xi(t')] =\delta(t-t')$, while $\gamma_x$ and $D_x$ are the damping and diffusion terms respectively, given by
\begin{equation}
\label{diffusion_damping}
    D_x=S_x(0),\qquad  m\gamma_x = \frac{\text{d}S_x(\omega)}{\text{d}\omega} \bigg\rvert_{\omega=0},
\end{equation}
with $S_x(\omega)$ the noise spectrum of the fluctuating force:
\begin{equation}
\label{noise_function}
    S_x(\omega) = F^2\int^\infty_{-\infty}\text{d}t\, e^{i\omega t} \left[\langle \hat{n}(t) \hat{n}(0)\rangle_x - \langle \hat{n}\rangle_x^2\right].
\end{equation}
In particular, since the oscillator's Wigner function is positive within the quasi-adiabatic approximation, it can be reconstructed by stochastically sampling the solutions of Eq.~\eqref{langevin_equation}. 

The dynamics of the clockwork depends qualitatively on the voltage bias. For zero or very low voltage bias, the leads behave effectively like an equilibrium bath for the oscillator, such that the fluctuation-dissipation relation $\beta D_x = 2m \gamma_x $ is approximately satisfied. The Langevin equation~\eqref{langevin_equation} then predicts a steady state for the oscillator that is almost thermal. Conversely, when the voltage exceeds a certain threshold value, electron tunnelling drives self-oscillations of the mechanical motion. In this regime, the damping rate $\gamma_x$ entering the Langevin equation~\eqref{langevin_equation} becomes negative for small $x$. As a result, the long-time state of the oscillator predicted by Eq.~\eqref{langevin_equation} is a limit cycle that oscillates with period $\sim 2\pi/\omega_0$~\cite{Wen2020,Culhane_2022}.

\subsection{Electrical current and current noise}
\label{sec:current_model}

Under a voltage bias, electrical current flows continuously through the system. Mechanical oscillations induce variations of this current that can be monitored to generate a series of ticks, as depicted in Fig.~\ref{fig:Schematic} and explained in more detail in Sec.~\ref{sec:corr_func}. A key assumption of our work is that the current signal is effectively classical, i.e. it can be read out without additional measurement backaction beyond that which is already contained in the damping and diffusion terms of Eq.~\eqref{langevin_equation}. This is justified in the quasi-adiabatic limit, because a very large number of electrons tunnel through the device within a given oscillation period. In practice, oscillations in the current can be continuously monitored by using a resonant tank circuit to filter frequency components near the relevant frequency, together with quantum-limited amplification~\cite{Wen2020,vigneau_probing_2023}.


To find how the oscillator motion modifies the current, we use the fact that $\hat{\rho}(x,p) \approx \mathcal{P}\hat{\rho}(x,p)$ to lowest order in small quantities~\cite{Culhane_2022}. The electronic state at long times is therefore given to lowest order by
\begin{equation}    \hat{\rho}_\text{el}\approx\int\text{d}x\int\text{d}p\, W(x,p)\hat{\rho}_x = \int\text{d}x\,P(x)\hat{\rho}_x.\label{el_state}
\end{equation}
where $W(x,p)$ is the steady-state Wigner function and $P(x)=\int\text{d}pW(x,p)$ is the corresponding marginal probability distribution for the oscillator's position. Equation~(\ref{el_state}) can be used to compute observables of the electronic system. In particular, the average of the electronic current operator, $\hat{I}$, is given to lowest order by
\begin{equation}
	\label{average_current}
    \langle\hat{I}\rangle = \text{tr}\left[\hat{I}\hat{\rho}_\text{el}\right] \approx \int\text{d}x\,P(x)\text{tr}\left[\hat{I}\hat{\rho}_x\right] = \int \text{d}x\,P(x)\braket{\hat{I}}_x.
\end{equation}
Here we have defined $\braket{\hat{I}}_x \equiv \tr \left[\hat{I}\hat{\rho}_x\right]$ as the conditional expectation value of the current operator, given the oscillator's position $x$.

While the ensemble-averaged current tends to the stationary value~\eqref{average_current} at long times, individual trajectories of the measured current will show periodic oscillations. This is because fluctuations of the oscillator position are transduced into current fluctuations, as characterised by the symmetrised autocorrelation function~\cite{clerk_introduction_2010}
\begin{align}
C(t) & = 	\tfrac{1}{2}\langle\{\hat{I}(t),\hat{I}(0)\}\rangle - \langle \hat{I}\rangle^2 \notag \\ &= \Re \left ( \tr \left[\text{e}^{i\hat{H}t}\hat{I}\text{e}^{-i\hat{H}t}\hat{I}\hat{\rho}\right]\right ) - \braket{\hat{I}}^2 \label{Curr_Corr_gen},
\end{align}
where all averages are taken with respect to the steady-state density matrix, $\hat{\rho}$. Let us define an auxiliary density operator
\begin{equation}
	\label{K_def}
    \hat{K}(t) = \text{e}^{-i\hat{H}t}\hat{I}\hat{\rho}\text{e}^{i\hat{H}t},
\end{equation}
such that the correlation function can be found from the real part of 
\begin{equation}\label{correlation_K}
	\langle\hat{I}(t)\hat{I}(0)\rangle = \tr \left[\hat{I}\hat{K}(t)\right] = \int\text{d}x\int\text{d}p \tr\left[\hat{I}\hat{K}(x,p,t)\right],
\end{equation}
where $\hat{K}(x,p,t)$ is the Wigner transform of of $\hat{K}(t)$, defined analogously to Eq.~\eqref{Wigner_transform}. Since $\hat{K}(t)$ obeys the Liouville-von Neumann equation
\begin{equation}
    \frac{\text{d}\hat{K}}{\text{d}t} = -i\left[\hat{H},\hat{K}\right],
\end{equation}
we have that $\hat{K}(x,p,t)$ obeys the same equation of motion as $\hat{\rho}(x,p))$: namely,
\begin{equation}
    \frac{\partial}{\partial t}\hat{K}(x,p,t) = (\mathcal{L}_x +\mathcal{H}+\mathcal{F})\hat{K}(x,p,t)\label{op_evo},
\end{equation}
where the superoperators on the right-hand side are defined in Eqs.~\eqref{L_super}--\eqref{F_super}. We can therefore approximate the solution to this equation following an adiabatic elimination procedure analogous to Sec.~\ref{sec:langevin}, but where now both $\mathcal{P}$ and $\mathcal{Q}$ projections are needed to reconstruct the correlation function. 

As shown in Appendix~\ref{app:corr_func}, this procedure yields the current autocorrelation function $C(t) = C_\mathcal{Q}(t) + C_\mathcal{P}(t)$, with 
\begin{align}
	\label{shot_noise}
	&C_\mathcal{Q}(t) =\mathbb{E}[\Delta_{x(0)}] \delta(t), \\ 
	 \label{current_correlation}
	& C_\mathcal{P}(t) = \mathbb{E}\left[\braket{\hat{I}}_{x(t)}\braket{\hat{I}}_{x(0)}\right] - \mathbb{E}\left[\braket{\hat{I}}_{x(0)}\right]^2.
\end{align}
In these expressions, all averages $\mathbb{E}[\bullet]$ are taken over oscillator trajectories $x(t)$ sampled from the Langevin equation~\eqref{langevin_equation}, with initial condition $x(0)$ distributed according to the steady-state position distribution, $P(x)$. Equation~\eqref{shot_noise} derives from the $\mathcal{Q}$ component of the correlation function, which is effectively delta-correlated in comparison to the slow timescales of the oscillator dynamics. This term represents the quantum shot noise of the current, $\Delta_x$, averaged over the steady-state position distribution. Equation~\eqref{current_correlation} is the $\mathcal{P}$ component of $C(t)$: it describes slow fluctuations of the current induced by stochastic variations of the oscillator position, $x(t)$.

For our model, both the conditional average current $\braket{I}_x$ and current noise $\Delta_x$ can be found using Landauer-B\"uttiker theory~\cite{ryndyk2016}, as
\begin{align}
	\label{current_LB}
	& \langle \hat{I}\rangle_x = \frac{1}{\pi}\int\dd E\, \tau_x(E)\left[f_L(E)-f_R(E)\right], \\
	& \Delta_x = \frac{2}{\pi}\int\dd E\tau_x(E)\left[f_L(E)(1-f_L(E))+f_R(E)(1-f_R(E))\right]\nonumber\\
    &\qquad+ \frac{2}{\pi}\int\dd E \tau_x(E)\left[1-\tau_x(E)\right]\left(f_L(E)-f_R(E)\right)^2\label{noise_LB}.
\end{align}
Here, $f_\alpha(E) = \left [ \ee^{\beta(E-\mu_\alpha)}+1\right ]^{-1}$ is the Fermi-Dirac distribution for lead $\alpha$, while $\tau_x(E)$ is the transmission function of the resonant-level model with the position-dependent energy $\epsilon-Fx$~\cite{ryndyk2016}:
\begin{equation}
    \label{transmission_function}
    \tau_x(E) = \frac{\kappa_L(E)\kappa_R(E)}{\left|E - \epsilon+Fx - \tilde{\chi}_L(
    E) - \tilde{\chi}_R(E)\right|^2,}
    \end{equation}
where $\kappa_\alpha(E)$ is the spectral density [Eq.~\eqref{spectral_density}] of lead $\alpha = L,R$ and the corrresponding self-energy is
\begin{equation}
    \tilde{\chi}_\alpha(E) = \fint \frac{\text{d}E'}{2\pi} \frac{\kappa_\alpha(E')}{E-E'} - \ii \frac{\kappa_\alpha(E)}{2},
\end{equation}
with $\fint$ a principal-value integral. 

The above equations are valid assuming coherent electron transport through the device, so that we can neglect Coulomb interactions and inelastic scattering. However, our conclusions are not strongly dependent on the choice of transport model. We have checked explicitly that qualitatively similar results are obtained using a sequential-tunnelling (i.e.~rate equation~\cite{Gurvitz1996}) description of transport.

\section{Results}

\subsection{\label{sec:corr_func}Analysis of current fluctuations}

\subsubsection{Current trajectories and timescale separation}

As described in Sec.~\ref{sec:current_model}, the mechanical motion induces changes in the measured current. Therefore, mechanical self-oscillations will lead to a periodic current signal, which can be used to measure the passage of time. The accuracy of such timekeeping is then limited by stochastic fluctuations in the current signal, as characterised by the autocorrelation function $C(t)$. The most relevant fluctuations for timekeeping are encoded in the slow part of the correlation function $C_\mathcal{P}(t)$ [Eq.~\eqref{current_correlation}] because this varies on the same timescale as the clockwork dynamics. The fast fluctuations described by $C_{\mathcal{Q}}(t)$ merely add a constant to the current noise spectrum, as discussed at the end of this section.

\begin{figure}

	\begin{center}

		\includegraphics[width=\columnwidth,height=6.5cm]{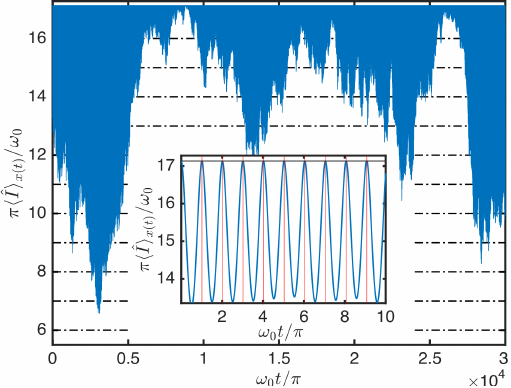}

		\caption{Portion of a single current trajectory above the self-oscillation voltage threshold, with $V=100\omega_0$. The inset shows the oscillations on a shorter time scale where the oscillations appear more sinusoidal. The red vertical lines indicate times when a tick is recorded. Parameters: $\lambda = 0.5\omega_0$, $\Gamma = 10\omega_0$, $\beta \omega_0 = 0.1$, $\omega_L = -\omega_R = 2.5\omega_0$, and $\delta = 5\omega_0$.\label{fig:Trajectory}}
		
	\end{center}
\end{figure}

To characterise the properties of the measured current signal, we solve the Langevin equation~\eqref{langevin_equation} using a stochastic Euler method and convert trajectories $x(t)$ into a corresponding current signal $\braket{\hat{I}}_{x(t)}$ via Eq.~\eqref{current_LB}. Fig.~\ref{fig:Trajectory} shows a sample current trajectory within the parameter regime where self-oscillations occur. We use the same parameters as in Ref.~\cite{Culhane_2022}, where the transition to self-sustained oscillations was found to occur at $V\approx 30\omega_0$. We observe near-periodic oscillations of the current on shorter timescales (Fig.~\ref{fig:Trajectory} inset), with a peak-to-peak amplitude that varies significantly over longer timescales (Fig.~\ref{fig:Trajectory} main panel). We also see that the oscillation frequency is approximately twice the natural frequency of the mechanical oscillator, and that the maximum oscillation amplitude has a fixed value that does not fluctuate. Both of these observations can be understood by the fact that the maximum current is obtained whenever the oscillator's position crosses $x=0$. A natural way to measure time is therefore to register a tick of the clock whenever the current reaches its maximum value, as illustrated in the inset of Fig.~\ref{fig:Trajectory}. Note that while the precise position $x$ at which the current is maximised depends on the specific choice of parameters, other parameters would yield qualitatively similar results.

\begin{figure}

	\begin{center}
		\includegraphics[width=\columnwidth,height=6.5cm]{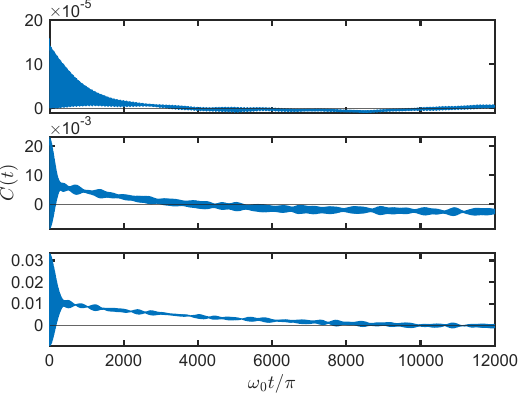}

		\caption{Auto-correlation function of the current for the same parameters as Fig.~\ref{fig:Trajectory} and three different voltages: $V=5\omega_0$ (top), $V=50\omega_0$ (middle), $V=100\omega_0$ (bottom). Two timescales emerge under high voltage bias: a fast-decaying component due to phase fluctuations and and a longer-lived component associated with amplitude noise. \label{fig:Two_Point_Corr}}

	\end{center}
\end{figure}

Figure~\ref{fig:Two_Point_Corr} shows the correlation function $C(t)$ for three different applied voltages, spanning the transition from thermal noise at low bias to self-sustained oscillations at high bias. At low voltage we observe decaying oscillations, as would be expected for the thermal fluctuations of an underdamped harmonic oscillator~\cite{risken_fokker-planck_1996}. 
Meanwhile, at voltages above the threshold for self-sustained oscillations, two distinct timescales can be discerned. The first, fast timescale governs the decay of oscillations, which results from phase diffusion on the limit cycle. The second, slow timescale dictates the asymptotic decay of the correlation function: for example, in the lower panel of Fig.~\ref{fig:Two_Point_Corr}, oscillations decay over a few hundred oscillation periods, while the correlation function itself does not decay completely until approximately $10^4$ periods have elapsed. This latter decay is monotonic rather than oscillatory and can, therefore, be understood as a consequence of long-lived amplitude correlations, as we now explain.

\subsubsection{Modelling amplitude and phase fluctuations}
\label{sec:phase_amplitude}

To elucidate the different roles of amplitude and phase fluctuations, in Appendix~\ref{app:OU_derivation} we derive an effective equation of motion describing small fluctuations of the oscillator trajectory away from the limit cycle. We express the results in polar coordinates, i.e.~$x = A\cos(\phi)$ and $\dot{x}/\omega_0 = - A\sin(\phi)$. An ideal limit cycle would have a constant amplitude $A = A_0$ and a linearly increasing phase $\phi = \omega_0 t$.  Taking into account fluctuations away from the limit cycle up to first order, we obtain an Ornstein-Uhlenbeck process describing amplitude variations, and a biased diffusion process describing the phase evolution, viz.
\begin{align}
	\label{amplitude_eqn}
   & \text{d}A = - \gamma_A(A-A_0)\text{d}t+\sqrt{D_A}\text{d}w_A, \\
    \label{phase_eqn}
   & \text{d}\phi = \omega_0\text{d}t +\sqrt{D_\phi}\text{d}w_\phi.
\end{align}
Here, $\gamma_A$ is the damping rate governing the regression of amplitude fluctuations back to the limit cycle $A_0$, $\dd w_A$ and $\dd w_\phi$ are independent Wiener increments describing amplitude and phase fluctuations, respectively, and  $D_A$ and $D_\phi$ are the corresponding diffusion coefficients. Our simplified model assumes that the limit-cycle amplitude is large and that the timescale of phase and amplitude diffusion is small compared to the oscillation frequency, which justifies approximating the amplitude and phase noise as independent even though they derive from the same Wiener process. The explicit expressions for the above parameters are cumbersome (see Appendix~\ref{app:OU_derivation} for details), but a straightforward numerical evaluation shows that $\gamma_A \gg D_\phi$ in the high-bias regime where Eqs.~\eqref{amplitude_eqn} and \eqref{phase_eqn} are valid. This indicates that small amplitude variations evolve significantly faster than phase decoherence, and therefore this model alone cannot explain the timescale separation evident in Fig.~\ref{fig:Two_Point_Corr}.

We thus turn to an alternative toy model that captures the large-scale amplitude variations seen in Fig.~\ref{fig:Trajectory}. Specifically,  the trajectory shown in Fig.~\ref{fig:Trajectory} exhibits rare periods where the oscillation amplitude reduces dramatically, so that the current remains close to its maximum value for a short time before returning to its previous oscillatory state. We model this behaviour as a bistable process, where the stochastic current switches between two states labelled by $i=1,2$, with corresponding currents $I_1(t)$ and $I_2(t)$.  This kind of bistable behaviour was predicted for electromechanical self-oscillations in Ref.~\cite{Usmani2007}, and electromechanical bistabilities have recently been observed experimentally~\cite{tabanera-bravo_stability_2022}.  We assume that these two states have different steady-state averages $\overline{I_i} = \mathbb{E}[I_i(t)]$, and that they are statistically independent from each other and from the switching process itself. Under these assumptions, we show in Appendix~\ref{app:Bistable} that the steady-state autocorrelation function takes the form
\begin{align}
	&C_{\mathcal{P}}(t) =\sum_{i=1}^2p_ip_{i|i}(t)C_i(t) + C_{1\leftrightarrow 2}(t), \label{Bi_Corr} \\ 
\label{C_switch}
	& C_{1\leftrightarrow 2}(t) = \sum_{i,j=1}^2 p_i \left[p_{j|i}(t)-p_j\right]\overline{I_i}\, \overline{I_j}, 
\end{align}
where $p_i$ is the steady-state probability to be in state $i$, $p_{j|i}(t)$ is the probability to transition $i\to j$ over a time $t$, and $C_i(t) = \mathbb{E}[I_i(t)I_i(0)] -\overline{I_i}^2$ is the auto-correlation function of $I_i(t)$. 

The first term in Eq.~\eqref{Bi_Corr} clearly quantifies correlations within each state. The second term, $C_{1\leftrightarrow 2}(t)$, is the contribution from the bistability itself: its time dependence arises only from the statistics of the switching process, while it depends on the current only through the conditional averages, $\overline{I_i}$. Therefore, the autocorrelation function will exhibit a separation of timescales whenever the switching process is much slower than phase and amplitude fluctuations of the current within each state. In Appendix~\ref{app:Bistable} we demonstrate this explicitly, showing that this simple bistable model recovers the features of the auto-correlation function above threshold in Fig.~\ref{fig:Two_Point_Corr}. Crucially, using the conservation of probability, $\sum_i p_i = \sum_j p_{j|i}(t) = 1$, one can show that $C_{1\leftrightarrow 2}(t) \propto \overline{I_1} - \overline{I_2}$. Therefore, it is important that the cycle-averaged current depends on the oscillation amplitude, which is indeed the case in our setup (cf. Fig.~\ref{fig:Trajectory}).

\subsubsection{Current noise power spectrum}
\label{sec:power_spectrum}

\begin{figure}
	\begin{center}
		
		\includegraphics[width=\columnwidth,height=7cm]{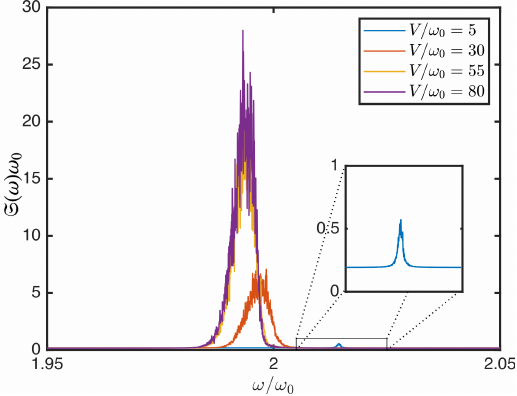}
		\caption{Power spectrum of the output current near the oscillation frequency $\omega \approx 2\omega_0$, for four different voltage biases; all other parameters are the same as Fig.~\ref{fig:Trajectory}. The frequency-independent contribution from $C_\mathcal{Q}(t)$ is small and only makes a significant contribution for smaller biases as shown in the inset.
			\label{fig:PowSpec}}
	\end{center}
\end{figure}

Finally, it is instructive to analyse the current noise in the Fourier domain via the power spectrum
\begin{equation}
	\label{power_spectrum}
	\mathfrak{S}(\omega) = \int \dd t\, \ee^{i \omega t} C(t).
\end{equation}
In addition to a flat white-noise contribution from the quantum shot-noise term $C_{\mathcal{Q}}(t)$, we observe two distinct peaks in the power spectrum at $\omega=0$ and $\omega\approx 2\omega_0$, reflecting the separation of timescales inherent to $C_{\mathcal{P}}(t)$. The height of both peaks grows with voltage bias, but for different reasons. The zero-frequency peak is associated with amplitude noise and its height scales with the lifetime of amplitude correlations (see Appendix~\ref{app:Bistable}), which becomes very large well above threshold. This growth of the zero-frequency noise is well-known in other non-equilibrium systems with long-lived metastable states~\cite{Usmani2007,kewming_diverging_2022}. Meanwhile, the peak in $\mathfrak{S}(\omega)$ at  $\omega\approx 2\omega_0$ is related to electromechanical oscillations, and is plotted in Fig.~\ref{fig:PowSpec}. The height of this peak scales with the oscillation amplitude, which is of course larger in the self-oscillating regime.

As the voltage increases, the position of the finite-frequency peak shifts slightly downwards in frequency, while its width becomes narrower. This implies that the ticks of the clock become more regular in time, while their frequency slightly decreases. As discussed in Sec.~\ref{sec:wtd}, these results indicate an increase in the clock's accuracy as the rate of dissipation increases, accompanied by a small decrease in the clock's resolution. This is consistent with previous studies of autonomous clocks ~\cite{Erker_2017,Schwarzhans2021,Pearson2021,meier_fundamental_2023}.

\subsection{\label{sec:wtd}Accuracy and resolution from the tick distribution}

We now turn to a more detailed analysis of the individual ticks of the clock. The tick distribution or waiting-time distribution (WTD) is defined as the probability, $W(\tau)$, of a time delay $\tau $ between two consecutive ticks. In simple models of autonomous clocks, each tick is assumed to be generated by an identical process, leading to the waiting times being independent and identically distributed (i.i.d.) random variables. Then, the WTD fully characterises the statistical properties of the time signal. 

In particular, the mean and variance of the WTD are 
\begin{align}
	\label{mean_WTD}
	 & \mu = \int_0^\infty \dd \tau\, \tau W(\tau),\\
	 \label{var_WTD}
	& \sigma^2 = \int_0^\infty \dd \tau\, (\tau-\mu)^2 W(\tau).
\end{align}
The clock's resolution $\nu$ and accuracy $N$ are then defined as~\cite{Erker_2017}
\begin{align}
	\label{resolution}
	& \nu = \mu^{-1},\\
	\label{accuracy}
	& N = \frac{\mu^2}{\sigma^2}. 
\end{align}
The resolution is simply the rate at which the ticks are generated, while the accuracy is the expected number of ticks before the clock's reading is off by one tick. To see this, suppose that $n$ ticks spaced by waiting times $\tau_1,\tau_2,\ldots \tau_n$ have been generated. The clock's time reading is $t_n = \sum_{i=1}^n \tau_i$, with average $\mathbb{E}[t_n] = n\mu$ and root-mean-square error $\Delta t_n = \sqrt{n}\sigma$ due to the assumption of i.i.d.~ticks. The accuracy is then the effective number of ticks $N$ such that $\Delta t_N = \mu$, which leads directly to Eq.~\eqref{accuracy}.

\begin{figure}
	\begin{center}

		\includegraphics[width=\columnwidth,height=7cm]{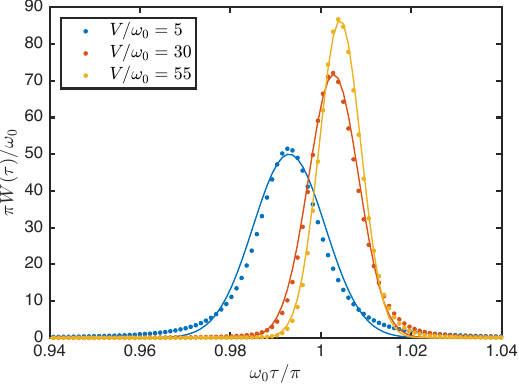}
		\caption{Waiting time distributions of the ticks for different voltage biases, with other parameters the same as Fig.~\ref{fig:Trajectory}. Dots are the data from the numeric calculations, the solid curves are the lines of best fit to the inverse Gaussian distribution~\eqref{Wald_distro}, with fitting parameters $\mu$ and $\sigma$. \label{fig:WTD}}
	
	\end{center}
\end{figure}

We construct the WTD for our model numerically as a normalised histogram of the waiting times between all pairs of consecutive ticks in the time series. The results are plotted in Fig.~\ref{fig:WTD} for three different voltages. Above threshold, we find that the WTD is well approximated by an inverse Gaussian (Wald) distribution
\begin{equation}
	\label{Wald_distro}
    W(\tau) = \sqrt{\frac{\mu^3}{2\pi\sigma^2 \tau^3}}\text{exp}\left(-\frac{\mu\left(\mu - \tau\right)^2}{2\sigma^2 \tau}\right).
\end{equation}
As shown in Ref.~\cite{Aminzare2019}, Eq.~\eqref{Wald_distro} can be derived by considering the distribution of the first-passage time $\tau$ for the oscillation phase on the limit cycle to complete one full $2\pi$ rotation. This result holds when the phase is described by a continuous-time biased random walk, as in Eq.~\eqref{phase_eqn}. The solid curves in Fig.~\ref{fig:WTD} show the best fits of the numerical WTDs to Eq.~\eqref{Wald_distro}, with $\mu$ and $\sigma$ as fitting parameters.  

\begin{figure}
	
	\begin{center}
		
		\includegraphics[width=\columnwidth]{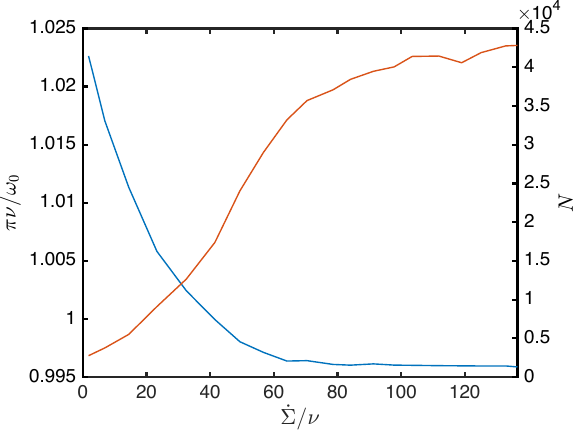}
		
		\caption{Accuracy (red line), and resolution (blue line) for increasing bias voltage, plotted against the entropy produced per tick of the clock. Parameters are the same as Fig.~\ref{fig:Trajectory}. \label{fig:Acc_Res}}
		
	\end{center}
\end{figure}

Notably, as the bias across the system is increased, the mean of the distribution increases and the variance of the distribution is decreased. This indicates that the clock's accuracy increases as the bias across the system is increased, while its resolution decreases slightly. Increasing the bias also increases the average rate of entropy production, which is proportional to the power dissipated in the circuit:
\begin{equation}
\label{entropy_production_rate}
    \dot{\Sigma} = \beta \langle 
    \hat{I}\rangle V.
\end{equation}
Figure~\ref{fig:Acc_Res} shows the accuracy and resolution of the electromechanical clock against the entropy produced per tick, $\Delta S_{\rm tick} = \dot{\Sigma}/\nu$. We observe an approximately linear relationship between accuracy and entropy production during the onset of self-sustained oscillations, showing that accurate timekeeping comes at the cost of increased dissipation. However, as the voltage is increased further, the growth of accuracy slows and begins to saturate to a value of around $4\text{x}10^4$: at this point, additional dissipation gives little further advantage to timekeeping. 

Meanwhile, the clock's resolution decreases and eventually saturates with increasing dissipation. However, the change in resolution is minimal for this model, which is unsurprising given that the tick period is fundamentally fixed by the physical oscillator frequency. The small decrease in resolution is probably due to effective non-linearities associated with the position dependence mean-field force $\propto  F\braket{\hat{n}}_x$ and damping rate $\gamma_x$, whose effect will change as the oscillator explores a larger region of phase space.

\subsection{\label{sect:allan_var} Allan variance}

The measures of accuracy and resolution considered in Sec.~\ref{sec:wtd} are only strictly applicable when the waiting times are i.i.d. random variables. The latter condition is equivalent to the statement that the total number of ticks is a renewal process~\cite{Cox_1967}. However, this assumption does not exactly hold because the limit cycle is mesoscopic in scale and subject to strong amplitude fluctuations. While the amplitude and phase noise can be approximated as independent for very small deviations from the limit cycle [see Sec.~\ref{sec:phase_amplitude}], this conclusion does not hold in general. Larger amplitude fluctuations will become correlated with phase noise because they both derive from the same underlying stochastic force induced by electron tunnelling. Therefore, differences in the state of the system after each tick may cause the next tick to be correlated with the previous ones.

Given that the ticks of our mesoscopic electromechanical clock are correlated, we need more a sophisticated measure of performance than the accuracy and resolution defined in Eqs.~\eqref{resolution} and~\eqref{accuracy}. We focus instead on the Allan variance~\cite{Allan_1966}, which measures the frequency stability of a stochastic process and is widely used to test the stability of sensors and clocks. The Allan variance quantifies the deviation of a measured clock signal $\theta$, from a second reference clock. For simplicity, we take this reference to be an ideal clock whose reading is the coordinate time, $t$, so that the measured deviation is
\begin{equation}
	\label{time_deviation}
	X(t) = \theta(t) - t.
\end{equation}
Then, $y(t) = \dd X/\dd t$ is the instantaneous frequency difference between the measured clock and the reference clock at a time $t$. This frequency deviation is estimated by observing the clock's reading over a finite integration time $T$, leading to the average fractional frequency difference
\begin{equation}
		\label{av_frac_freq_diff}
	\bar{y}(t,T) =\frac{1}{T}\int^T_0\text{d}t'\, y(t+t') = \frac{X(t+T)-X(t)}{T}. 
\end{equation}
The Allan variance is then proportional to the mean-square deviation of successive average fractional frequency differences
\begin{equation}
\label{Allan_variance_def}
	\sigma^2_y(T) = \frac{1}{2}\mathbb{E}\left\{\left[\bar{y}(t+T,T) - \bar{y}(t,T)\right]^2 \right\},
\end{equation}
thus quantifying how instability in the average frequency scales with integration time. In the case of deterministic signals, where the frequency remains constant, the anticipated frequency deviation is zero. Consequently, a clock that closely adheres to deterministic behaviour --- and is, therefore, more accurate --- would exhibit a smaller Allan variance. 

We compute the Allan variance for our electromechanical clock by partitioning the stochastic current trajectories into non-overlapping regions of equal duration $T$. The measured deviation pertaining to the $n^{\rm th}$ region, $X_n = X(n\tau)$, is evaluated by counting the number of registered ticks up to time $t=n\tau$. Finally, the Allan variance is computed as
\begin{equation}
	\label{allan_second_def}
	\sigma_y^2(T) = \frac{1}{2T^2}\mathbb{E}\left[ \left(X_{n-1} - 2X_{n} + X_{n+1}\right)^2\right].
\end{equation}
The behaviour of the Allan variance as a function of $T$ may be sensitive to the precise choice of $T$ values --- we have repeated our calculations for different values of $T$ to ensure that the results are robust.

\begin{figure}
	\begin{center}
		
		\includegraphics[width=\columnwidth,height=6.5cm]{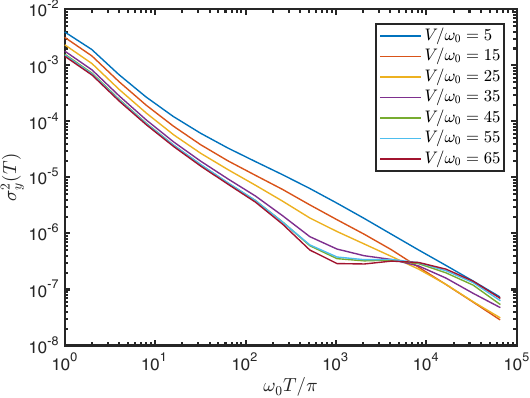}
		
		\caption{Allan variance as a function of integration time $T$ for various different voltage biases. Parameters are the same as Fig.~\ref{fig:Trajectory}. \label{fig:Allan}}
		
	\end{center}
\end{figure}

The results for the Allan variance are plotted in Fig.~\ref{fig:Allan}, for a range of different voltages. In general, the Allan variance decreases systematically with $T$: evidently, integrating the clock signal over a longer time yields greater precision. This dynamic highlights the trade-off between the clock's timekeeping accuracy and resolution~\cite{meier_fundamental_2023}. Note that here $\sigma_y^2$ decreases indefinitely with $T$, whereas real experiments usually find that the Allan variance begins to grow beyond a certain integration time. This is typically due to the influence of low-frequency fluctuations, e.g.~due to temperature variations or the gradual degradation of the apparatus over time, which are absent in our theoretical model. 

At short and intermediate timescales, we observe from Fig.~\ref{fig:Allan} that higher voltage leads to a smaller values of $\sigma_y^2$, demonstrating a quantitative link between accuracy and the rate of energy dissipation. It follows that the relationship between timekeeping precision and entropy production extends over multiple ticks of the clock, beyond the simple accuracy measure of Eq.~\eqref{accuracy} that only accounts for the single-tick WTD. At long times, all curves in Fig.~\ref{fig:Allan} tend to the approximate scaling $\sigma_y^2 \sim T^{-1}$. This is consistent with known results for the long-time Allan variance for a renewal process (see Appendix~\ref{app:allan} or Ref.~\cite{silva_ticking_2023}):
\begin{equation}
	\label{renewal_allan_variance}
	\sigma_y^2 \to \frac{\mu}{N T},
\end{equation}
which is fully determined by the clock's accuracy $N$ defined in Eq.~\eqref{accuracy}. A similar $1/T$ scaling is observed in Fig.~\ref{fig:Allan} across essentially all times for voltages below the threshold for self-oscillations. 

Interestingly, however, the Allan variance initially decreases slightly faster than $T^{-1}$ for voltages above threshold. This trend continues up to an intermediate timescale on the order of a few thousand ticks, after which the slope of $\sigma_y^2(T)$ changes dramatically and the advantage in timekeeping precision due to self-oscillations is eventually lost altogether. The non-trivial behaviour of the Allan variance as a function of $T$ indicates that temporal correlations play a significant role in determining the clock's perfomance over multiple ticks.

\subsection{Quantifying correlations between ticks}
\label{sec:tick_correlations}

In this section we turn to explicitly quantifying correlations between the ticks of the clock, which are responsible for the faster decrease of $\sigma_y^2(T)$ above threshold. We consider two different measures of statistical correlations. First, we consider the clock reading after $n$ ticks, $t_n=\sum_{i=1}^n \tau_i$, which is the sum of $n$ consecutive waiting times $\tau_1,\tau_2,\ldots, \tau_n$. Let $P_n(t_n)$ be the corresponding probability distribution. If the ticks were generated by a renewal process, so that the individual waiting times were i.i.d., this distribution would be given by the $n$-fold convolution of single-tick WTDs, i.e.
\begin{align}
	\label{Conv_WTD}
	\left [\Conv^n W\right ](t_n) =  \int_0^{t_n} & \text{d}t_1 \int_{t_1}^{t_n}\dd t_2  \cdots \int_{t_{n-2}}^{t_n}  \text{d}t_{n-1}  W(t_1) \times \notag  \\ 
	& W(t_2-t_1)  \times  \cdots  \times W(t_n-t_{n-1}).
\end{align}
We use the relative entropy (Kullback-Leibler divergence) to measure the distance between $P_n(t_n)$ and the i.i.d.~result in Eq.~\eqref{Conv_WTD}:
\begin{equation}
	\label{relative_entropy}
	D_{\rm KL}(P_n ||\Conv^n W) = \int_0^\infty \text{d}t\, P_n(t)\ln \left(\frac{P_n(t)}{\left [\displaystyle\Conv^n W\right ](t)}\right).
\end{equation}
Since we are considering a stationary process, if the waiting times are not i.i.d.~then they are not independent. Equation~\eqref{relative_entropy} can thus be understood as a measure of the total statistical correlations among all $n$ ticks. 

\begin{figure}
	\begin{center}
		\includegraphics[width=\columnwidth]{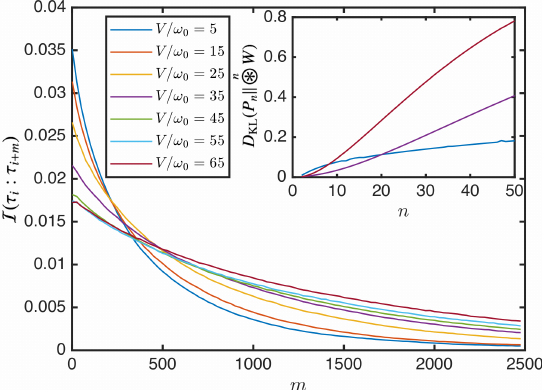}
		\caption{The main panel shows the mutual information [Eq.~\eqref{mutual_info}] between a tick and the $m^{\rm th}$ subsequent tick, i.e.~the pairwise correlations between the waiting times $\tau_i$ and $\tau_{i+m}$. The inset shows the relative entropy [Eq.~\eqref{relative_entropy}] between the distribution of the $n$-tick clock reading $P(t_n)$ and the $n$-fold convolution of single-tick WTD, which measures the total correlations among all $n$ ticks. Parameters are the same as Fig.~\ref{fig:Trajectory}. \label{fig:Rel_Ent}}
	\end{center}
\end{figure}

The inset of Fig.~\ref{fig:Rel_Ent} demonstrates that the ticks are indeed correlated: the relative entropy grows as a function of $n$. Moreover, this growth is faster for higher voltages, indicating that correlations between ticks become more significant in the self-oscillation regime. Since these correlations should have a finite range in time, we expect the total correlations to eventually saturate as $n$ increases. Unfortunately, we are unable to reliably extend our calculation of $D_{\rm KL}(P_n ||\Conv^n W) $ to higher values of $n$: the results become very noisy due to the limited quantity of available numerical data and the relative complexity of the $n$-fold convolution.

To estimate the timescale over which correlations decay, we therefore focus on a simpler measure of correlations: the mutual information between pairs of ticks that are $m$ ticks apart. This is defined as
\begin{equation}
	\label{mutual_info}
	\mathcal{I}(\tau_i: \tau_{i+m}) = 2H[W(\tau_i)] - H[W(\tau_i,\tau_{i+m})],
\end{equation}
where $W(\tau_i,\tau_{i+m})$ is the joint distribution of the waiting times $\tau_i$ and $\tau_{i+m}$, $W(\tau_i)$ is the WTD (identical for $\tau_i$ and $\tau_{i+m}$) and $H[W(\tau)] = -\int\dd \tau W(\tau)\ln W(\tau)$ denotes the Shannon entropy. Unlike the relative entropy, which measures the total correlations among all ticks, Eq.~\eqref{mutual_info} quantifies only correlations between specific pairs of ticks and can be computed reliably over long timescales using our numerical data. The mutual information is plotted in the main panel of Fig.~\ref{fig:Rel_Ent}. This plot demonstrates that indeed the correlations are short-ranged in time: $\mathcal{I}(\tau_i: \tau_{i+m})$ is largest for small $m$ and at lower voltages. However, especially at voltages above threshold, these correlations persist over long timescales on the order of several thousand ticks. This is the same timescale at which the non-monotonic behaviour of the Allan variance in Fig.~\ref{fig:Allan} appears.


We thus conclude that $\sigma_y^2$ decreases more rapidly above threshold because of temporal correlations between the ticks. These correlations provide an advantage in timekeeping precision, so long as the integration time $T$ is not much greater than the correlation time. By contrast, when $T$ far exceeds the range of temporal correlations, the integrated signals in two neighbouring time periods become effectively independent and the behaviour of the Allan variance reduces to that of a renewal process, i.e. $\sigma_y^2 \sim T^{-1}$. 

These correlations are likely dictated by several factors acting in tandem. The timescale over which they persist appears to be far longer than the timescale of pure phase decoherence apparent in Fig.~\ref{fig:Two_Point_Corr}, indicating that amplitude fluctuations also play a role. This can be understood from the fact that amplitude and phase fluctuations are generally not independent, meaning that a variation in amplitude after one cycle can influence the time taken for the next cycle to be completed. The timescale over which tick correlations persist will then be determined by a combination of phase decoherence, amplitude fluctuations near the limit cycle, and large-scale amplitude variations due to bistability in the current signal.

\section{\label{sect:disc}Conclusions and outlook}

In this work, we have introduced a self-contained model of an autonomous clock, where the clockwork is an electromechanical self-oscillator and the ticks of the clock are read out from the electrical current passing through the system. A peculiar and novel feature of our setup is that the clockwork is driven by the same irreversible process that is used for tick readout --- the tunnelling of electrons through the device. We have confirmed that the clock's accuracy increases with the rate of entropy production, as the current drives the oscillator into a limit cycle with long-lived phase coherence. We have identified that both phase and amplitude fluctuations play an important role in determining the statistical properties of the measured current. 

A key finding of our work is that the ticks of the clock are generally correlated with each other and that these correlations can improve the clock's accuracy, as quantified by the Allan variance. It remains an open problem to understand under what conditions such correlations can improve clock accuracy in general. We note that temporal correlations have recently been considered in the context of (classical) parameter estimation, where they may either help or hinder depending on their nature~\cite{radaelli_fisher_2023}. Yet far more work is needed to unify the theory of quantum ticking clocks with the standard framework of quantum parameter estimation, especially beyond the asymptotic regime of infinitely many samples (or in this case, ticks). 

Our model is strongly motivated by modern electromechanical platforms, meaning that our predictions could be probed directly in state-of-the-art experiments. In the present work, we have limited ourselves to the quasi-adiabatic regime, which is often the most experimentally relevant one and is also the simplest regime to treat theoretically. However, the quasi-adiabatic regime is very far from thermodynamically optimal, since many electrons tunnel through the device per tick of the clock. This regime is also sub-optimal from the perspective of readout, since the mechanical oscillations are ``oversampled'' by the tunnelling events~\cite{meier_fundamental_2023}. It is thus important to explore the complementary regime where each oscillation is sampled by only a handful of electrons, approaching the lower sensitivity limit of electrical readout and entailing thus a far smaller rate of dissipation. In this regime, fluctuations in the tunnelling time of individual electrons (e.g.~see Ref.~\cite{Altaie2022} and references therein) may become relevant for the clock's performance. At the level of theory and modelling, going beyond the quasi-adiabatic limit requires a non-perturbative methodology capable of tackling situations where the electromechanical coupling and the tunnelling rates are comparable. In future work, we plan to apply the mesoscopic-leads/extended-reservoir approach~\cite{Brenes_2020,lacerda_quantum_2023} to address this problem.

It is interesting to note that our model predicts clock accuracies (Fig.~\ref{fig:Acc_Res}) that significantly surpass thermodynamic performance bounds that apply to classical clocks with a discrete state space~\cite{Barato_2016}. In particular, the thermodynamic uncertainty relation~\cite{barato_thermodynamic_2015,gingrich_dissipation_2016} limits the accuracy of such clocks to one half of the entropy production per tick, i.e.~$N\leq \dot{\Sigma}/2\nu$ (assuming the ticks are generated by a renewal process)
. The fact that our electromechanical clock can surpass these bounds is unsurprising: the dynamics of our clock is governed by an underdamped classical Langevin equation, which does not necessarily obey these bounds~\cite{pietzonka_classical_2022,Van_Vu_2019}. Nevertheless, the approximations inherent to our model do neglect some potentially relevant contributions, which lead us to overestimate the accuracy and underestimate the entropy production. 

First, our definition of the ticks is based on the quasi-adiabatic current fluctuations quantified by $C_{\mathcal{P}}(t)$. We have neglected the shot-noise contribution due to the discrete nature of the electronic charge, i.e.~$C_{\mathcal{Q}}(t)$. This additional source of white noise would partially obscure the precise time at which ticks occur. To improve tick identification in the presence of such noise, one could redefine the tick time to occur when the time derivative of the current, rather than the current itself, is at its maximum~\cite{Pearson2021}. Second, reading out relatively small currents would require amplification in practice, and we have ignored the additional dissipation associated with this measurement step. We note that the thermodynamic cost of the measurement has also been neglected in previous work on autonomous clocks~\cite{Erker_2017,Schwarzhans2021,woods_autonomous_2021,silva_ticking_2023}. Understanding how this additional source of dissipation enters the balance of entropy production for timekeeping is an important open question for future work.

Finally, it would be interesting to explore the thermodynamics of timekeeping for other kinds of self-oscillators in the quantum regime. Conditions for self-oscillations to emerge from open quantum dynamics have only recently been established~\cite{chan_limit-cycle_2015,iemini_boundary_2018,buca_non-stationary_2019,guarnieri_time_2022}, within the broader context of research on broken time-translation symmetry~\cite{wilczek_quantum_2012,sacha_time_2018,zaletel_colloquium_2023}. However, a general recipe to exploit such symmetry breaking for accurate timekeeping remains lacking. 

\begin{acknowledgments}
We are grateful to N. Ares, J.~Dexter, M.~Huber, F.~Meier, P.~P.~Potts and R.~Silva for informative discussions.  This work was funded by the Irish Research Council, the European Research Council Starting Grant ODYSSEY (Grant Agreement No.~758403), the EPSRC-SFI joint project QuamNESS. This project is co-funded by the European Union (Quantum Flagship project ASPECTS, Grant Agreement No. 101080167). Views and opinions expressed are however those of the authors only and do not necessarily reflect those of the European Union, REA or UKRI. Neither the European Union nor UKRI can be held responsible for them. J.G. is supported by a SFI-Royal Society University Research Fellowship. M.J.K. acknowledges the financial support from a Marie Sk\l odwoska-Curie Fellowship (Grant Agreement No.~101065974). A.S. acknowledges financial support from PNRR MUR project PE0000023-NQSTI and Quantera project SuperLink. M.T.M. is supported by a Royal Society-Science Foundation Ireland University Research Fellowship (URF\textbackslash R1\textbackslash 221571). We acknowledge the Irish Centre for High End Computing for the provision of computational facilities. Some calculations were performed on the Kelvin cluster maintained by the Trinity Centre for High Performance Computing. This cluster was funded through grants from the Higher Education Authority, through its PRTLI program.
\end{acknowledgments}

\appendix

\section{Derivation of the electronic correlation function\label{app:corr_func}}

In this appendix, we derive the expression~\eqref{current_correlation} for the electronic correlation function. The starting point is the auxiliary density operator defined in Eq.~\eqref{K_def}, i.e.
\begin{equation}
	\label{K_def_app}
	\hat{K}(t) = \text{e}^{-i\hat{H}t}\hat{I}\hat{\rho}\text{e}^{i\hat{H}t}.
\end{equation}
Taking the Wigner transform as in Eq.~\eqref{Wigner_transform}, we have the equation of motion 
\begin{equation}
	\frac{\partial}{\partial t}\hat{K}(x,p,t) = (\mathcal{L}_x +\mathcal{H}+\mathcal{F})\hat{K}(x,p,t)\label{op_evo_app},
\end{equation}
with the superoperators given by Eqs.~\eqref{L_super}--\eqref{F_super}. The projection operator $\mathcal{P}$ defined in Eq.~\eqref{projection}, and its orthogonal complement $\mathcal{Q} = 1-\mathcal{P}$, obey the following properties~\cite{Culhane_2022}
\begin{subequations}
	\label{projector_properties}
	\begin{align}
		\label{projector_defining_property}
		\mathcal{P}^2 & = \mathcal{P}, \\
		\label{projector_kernel}
		\mathcal{P}\mathcal{L}_x & = 0 = \mathcal{L}_x \mathcal{P} , \\
		\label{projector_no_mean}
		\mathcal{P}\mathcal{F}\mathcal{P} & = 0,\\
		\label{Ham_proj_commutator}
		[\mathcal{P},\mathcal{H}] & = \frac{p}{m} \pd{\r_x}{x}\tr[\bullet], \\
		\label{proj_comm_vanish}
		\mathcal{P}[\mathcal{P},\mathcal{H}] &= 0.
	\end{align}
\end{subequations}
As in Ref.~\cite{Culhane_2022}, we assume that $\mathcal{L}_x\gg \mathcal{H},\mathcal{F}$, which follows from the quasi-adiabatic conditions~\eqref{adiabatic_condition}. Then the electronic degrees of freedom can be assumed to relax quickly to the stationary state of $\mathcal{L}_x$, while $\mathcal{H}$ and $\mathcal{F}$ can be treated perturbatively.

To proceed, we define the compact notation $\hat{r}(t) = \mathcal{P}\hat{K}(x,p,t)$ and $\hat{q}(t) = \mathcal{Q}\hat{K}(x,p,t)$, suppressing phase-space arguments for concision. Inserting appropriate factors of the identity $1=\mathcal{Q}+\mathcal{P}$ on both sides of Eq.~\eqref{op_evo_app}, and using the properties~\eqref{projector_properties}, one has 
\begin{align}
	\label{P_eqn}
	\pd{\hat{r}}{t} & = \mathcal{P}\mathcal{H}\hat{r} +\mathcal{P}\left (\mathcal{H} + \mathcal{F}\right )\hat{q}, \\
	\label{Q_eqn}
	\pd{\hat{q}}{t} & = \mathcal{L}_x \hat{q} + \mathcal{Q}\left (\mathcal{H} + \mathcal{F}\right )\hat{q} + \mathcal{Q}(\mathcal{H} + \mathcal{F})\hat{r}.
\end{align}
Formally integrating Eq.~\eqref{Q_eqn} yields \begin{align}\label{Q_soln}
	& \hat{q}(t) = \mathcal{G}(t)\hat{q}(0) + \int_0^t\dd t' \mathcal{G}(t-t') \mathcal{Q}\left (\mathcal{H + \mathcal{F}}\right )\hat{r}(t'), \\
	\label{G_def}
	& \mathcal{G}(t) = \exp \left \lbrace\left [ \mathcal{L}_x + \mathcal{Q} (\mathcal{H}+\mathcal{F})\right]t \right \rbrace.
\end{align}
Substituting this back into Eq.~\eqref{P_eqn}, we obtain the closed equation of motion
\begin{align}
	\label{PK_equation}
	\pd{\hat{r}}{t} & = \mathcal{P}\mathcal{H}\hat{r}(t) + \int_0^t\dd t'  \mathcal{P}\left (\mathcal{H}+\mathcal{F} \right )\mathcal{G}(t-t') \mathcal{Q}\left (\mathcal{H}+\mathcal{F} \right )\hat{r}(t') \notag \\
	& \quad  + \mathcal{P}
	\left (\mathcal{H}+\mathcal{F}\right ) \mathcal{G}(t)\hat{q}(0).
\end{align}
The first two terms on the right-hand side (RHS) of Eq.~\eqref{PK_equation} are identical to Eq.~(A14) in the Supplemental Material of Ref.~\cite{Culhane_2022}, and can be treated by identical manipulations. In brief, we make the following approximations: (i)~since $ \mathcal{H},\mathcal{F} \ll \mathcal{L}_x$ we write $\mathcal{G}(t) \approx \ee^{\mathcal{L}_x t}$ to lowest non-trivial order in small quantities (Born approximation), (ii)~assuming that the integrand decays rapidly as a function of $t'$, we approximate $\hat{r}(t-t') \approx \hat{r}(t)$ and extend the upper integration limit to infinity (Markov approximation). After tracing over the electronic degrees of freedom, therefore, the first two terms on the RHS of Eq.~\eqref{PK_equation} reduce to the Fokker-Planck generator
\begin{align}
\mathcal{L}_{\rm FP}K   = & \frac{\partial}{\partial p}  \left (m \omega_0^2 x - F\braket{\hat{n}}_x + \gamma_x p  \right )K    - \frac{p}{m} \frac{\partial K}{\partial x}  + \frac{D_x }{2} \frac{\partial^2K}{\partial p^2},
\end{align}
where $K(x,p,t) = \tr \hat{K}(x,p,t)$. The same Fokker-Planck operator dictates the evolution of the Wigner function, $\partial W/\partial t = \mathcal{L}_{\rm FP} W$, which can therefore be reconstructed by sampling the equivalent Langevin equation~\eqref{langevin_equation}. However, unlike the Wigner function, the equation for $K(x,p,t)$ has an additional inhomogenity deriving from the third term on the RHS of Eq.~\eqref{PK_equation}, i.e.
\begin{equation}\label{PK_FokkerPlanck}
	\pd{K}{t} = \mathcal{L}_{\rm FP} K + \tr\left [ \mathcal{P}
	\left (\mathcal{H}+\mathcal{F}\right ) \mathcal{G}(t)\hat{q}(0)\right ].
\end{equation}
This equation has the solution
\begin{equation}\label{K_solution}
	K(t) = \ee^{\mathcal{L}_{\rm FP} t} K(0) + \int_0^t\dd t' \, \ee^{\mathcal{L}_{\rm FP} (t-t')} \tr\left [ \mathcal{P}
	\left (\mathcal{H}+\mathcal{F}\right ) \mathcal{G}(t')\hat{q}(0)\right ].
\end{equation}
Since $\mathcal{P}\hat{K}(x,p,t) = K(x,p,t) \hat{\rho}_x$ and $\mathcal{P}\hat{K}(x,p,t)$ is given by Eq.~\eqref{Q_soln}, we have a formal solution for both components of $\hat{K}(x,p,t)$.

Now, since we are considering the steady-state current correlation function, the initial condition of Eq.~\eqref{op_evo_app} is 
\begin{equation}\label{K_initial_condition}
	\hat{K}(x,p,0) = \hat{I} \hat{\rho}(x,p) \approx  \hat{I} \hat{\rho}_x W(x,p),
\end{equation}
to lowest order in small quantities, where $\hat{\rho}(x,p)$ and $W(x,p)$ correspond to the stationary state. Applying the projection operators, we get at lowest order
\begin{align}\label{P_K}
	& \mathcal{P}\hat{K}(x,p,0) \approx \hat{\rho}_x \braket{\hat{I}}_x W(x,p) , \\
	\label{Q_K}
	&\mathcal{Q}\hat{K}(x,p,0) \approx \left (\hat{I}-\braket{\hat{I}}_x\right ) \hat{\rho}_xW(x,p).
\end{align}
To the same order, the second term on the right-hand side of both Eqs.~\eqref{Q_soln} and \eqref{K_solution} can be neglected and we can write $\mathcal{G}(t) \approx \ee^{\mathcal{L}_x t}$. Therefore, we finally obtain the approximate solution
\begin{align}
	\label{K_final_sol}
	\hat{K}(x,p,t) & = \mathcal{P} \hat{K}(x,p,t)  + \mathcal{Q} \hat{K}(x,p,t) \notag \\
	& \approx  \hat{\rho}_x \ee^{\mathcal{L}_{\rm FP} t} \!\left [\braket{ W(x,p)} \right ]  + \ee^{\mathcal{L}_x t} \!\left [\left (\hat{I} - \braket{\hat{I}}_x\right ) \hat{\rho}_x\right ] \! W(x,p).
\end{align}
Plugging this into Eq.~\eqref{correlation_K} yields
\begin{align}\label{I_intermediate}
	\braket{\hat{I}(t) \hat{I}(0)} & = \int \dd x\int \dd p \, \braket{\hat{I}}_x \ee^{\mathcal{L}_{\rm FP} t} \!\left [\braket{\hat{I}}_x W(x,p) \right ] \notag \\
	& \quad + \int \dd x\int \dd p \, W(x,p) \left [ \braket{\ee^{\mathcal{L}^\dagger_xt}(\hat{I})\hat{I}}_x - \braket{\hat{I}}^2_x \right ].
\end{align}

The first term in Eq.~\eqref{I_intermediate} can be interpreted as an autocorrelation function of the stochastic variable $\braket{\hat{I}}_{x(t)}$ with respect to the Langevin evolution of the oscillator. To see this, note that $ \ee^{\mathcal{L}_{\rm FP} t} \!\left [\braket{\hat{I}}_x W(x,p) \right ]$ is the solution of the Fokker-Planck equation with initial condition $\braket{\hat{I}}_x W(x,p)$. We define the Fokker-Planck propagator $\Pi(x',p',t'| x,p,t)$ as the conditional probability to find the oscillator at the phase space point $(x',p')$ at time $t'$ given that it was at $(x,p)$ at time $t$. We may therefore write the first term in Eq.~\eqref{I_intermediate} as
\begin{widetext}
	\begin{align}\label{FP_propagator}
		\int \dd x\int \dd p \, \braket{\hat{I}}_x \ee^{\mathcal{L}_{\rm FP} t} \!\left [\braket{\hat{I}}_x W(x,p) \right ] & = \int \dd x' \int\dd p'\int \dd x\int \dd p \, \braket{\hat{I}}_x  \Pi(x,p,t | x',p',0) \braket{\hat{I}}_{x'} W(x',p') \notag \\
		& = \int \dd x' \int\dd p'\int \dd x\int \dd p \, \braket{\hat{I}}_x  \braket{\hat{I}}_{x'} P(x,p,t ; x',p',0) \notag \\
		& \equiv \mathbb{E}\left [\braket{\hat{I}}_{x(t)}\braket{\hat{I}}_{x(0)} \right ], 
	\end{align}
\end{widetext}
where $P(x,p,t ; x',p',0) =  \Pi(x,p,t | x',p',0)W(x',p') $ is the joint probability to find the system initially (time $t=0$) at $(x',p')$ and finally (time $t$) at $(x,p)$, while $\mathbb{E}[\bullet]$ denotes an average over the joint distribution which can be computed by sampling Langevin trajectories.

The second term in Eq.~\eqref{I_intermediate} represents the instantaneous current fluctuations conditioned on the oscillator position $x$. Within the quasi-adiabatic limit, this correlation decays very quickly compared to the slow timescales of the oscillator. Thus, on timescales relevant for the oscillator, the real part of this correlation function can by approximated by a Dirac delta function:
\begin{equation}
\Re	\left (\braket{\ee^{\mathcal{L}^\dagger_xt}(\hat{I})\hat{I}}_x - \braket{\hat{I}}^2_x\right )  \approx \Delta_x \delta(t),
\end{equation}
where $\Delta_x$ is the steady-state current noise conditioned on $x$, i.e. 
\begin{equation}
	\label{current_noise_def}
	\Delta_x = \int_{-\infty}^\infty \dd t\, \Re	\left (\braket{\ee^{\mathcal{L}^\dagger_xt}(\hat{I})\hat{I}}_x - \braket{\hat{I}}^2_x\right ).
\end{equation}
Putting everything together, we obtain Eq.~\eqref{current_correlation}. Finally, we note that the above derivation holds for an arbitrary operator $\hat{I}$ whose support is restricted to the electronic system in the Schr\"odinger picture (otherwise Eq.~\eqref{K_initial_condition} does not hold).

\section{\label{app:OU_derivation} Modelling small amplitude and phase fluctuations near the limit cycle}

In this appendix, we derive a simple description of phase and amplitude fluctuations on the oscillator's limit cycle. Our starting point is the Langevin equation~\eqref{langevin_equation} governing the evolution of the oscillator. We first transform the equation to polar coordinates. Using the substitution $z\omega_0=v$ and converting the system to polar coordinates, $x=A\text{cos}\phi$, $z=A\text{sin}\phi$ the equations of motion become
\begin{align}
    \text{d}A &= \text{cos}\phi\text{d}x + \text{sin}\phi\text{d}z + \frac{\text{cos}^2\phi}{2A} \text{d}z^2\label{Amp_Eq},\\
    \text{d}\phi = &-\frac{\text{sin}\phi}{A}\text{d}x + \frac{\text{cos}\phi}{A}\text{d}z -\frac{\text{cos}\phi\text{sin}\phi}{2A^2}\text{d}z^2.\label{Phase_Eq}
\end{align}
Here we have used It\^{o}'s lemma, where $\text{d}z^2=\frac{D_x}{\omega_0^2}\text{d}t$, $\text{d}x^2=0$ and $\text{d}x\text{d}z=0$. Focusing first on the amplitude equation, the equation of motion is given as
\begin{align}
    \text{d}A = -m\gamma_{(A,\phi)}&A\text{sin}^2(\phi)\text{d}t + \frac{F}{\omega_0}\langle n\rangle_{(A,\phi)}\text{sin}(\phi)\text{d}t\nonumber\\
    &+\frac{\sqrt{D_{(A,\phi)}}}{w_0}\text{sin}(\phi)\text{d}w + \frac{\text{cos}^2(\phi)}{2A}\frac{D_{(A,\phi)}}{\omega_0^2}\text{d}t.\label{Amp_diff_eq}
\end{align}
Here $dw$ is a Wiener increment, and the subscript $(A,\phi)$ is used to highlight dependence on both amplitude and the phase. Assuming the amplitude change over a single cycle is small, we can replace the deterministic terms with their mean over a single oscillation and the stochastic term with its RMS over a single oscillation;
\begin{widetext}
\begin{align}
      \text{d}A & = -\frac{A}{2\pi}\oint\Big(\gamma_{(A,\phi)}\text{sin}^2(\phi)\Big)\text{d}\phi\text{d}t + \frac{F}{2\pi\omega_0}\oint\Big(\langle n\rangle_{(A,\phi)}\text{sin}(\phi)\Big)\text{d}\phi\text{d}t  +\frac{1}{\sqrt{2\pi}\omega_0}\sqrt{\oint\Big(\sqrt{D_{(A,\phi)}}\text{sin}(\phi)\Big)^2\text{d}\phi}\text{d}w_A \notag\\
& \quad  + \frac{1}{4\pi A\omega_0^2}\oint\Big(D_{(A,\phi)}\text{cos}^2(\phi)\Big)\text{d}\phi\text{d}t\label{A_diff_approx}.
\end{align}
\end{widetext}
Here $\dd w_A$ is an effective coarse-grained Wiener increment. Due to symmetry in the $\langle\hat{n}\rangle_x$ term, its average over a single cycle will be zero. We can find the limit cycle of the system by taking the expectation value of both sides of the equation and setting $\frac{\Delta\mathbb{E}\left[A\right]}{\Delta t} =0$. The limit cycle is given by the solution of the transcendental equation
\begin{equation}
    1 = \frac{1}{2\mathbb{E}[A]^2\omega_0^2}\frac{\mathbb{E}\left[\oint\text{cos}^2(\phi)D_{(A,\phi)}\text{d}\phi\right]}{\mathbb{E}\left[\oint\text{sin}^2(\phi)\gamma_{(A,\phi)}\text{d}\phi\right]}.
\end{equation}
We can also truncate Eq.~\ref{A_diff_approx} to lowest order for the dissipation and the fluctuation terms around the limit cycle. Using the substitutions
\begin{align}
     \label{gamma_A}
     &\gamma_A =\frac{\dd}{\dd A}\Bigg(-\frac{A}{2\pi}\oint\Big(\gamma_{(A,\phi)}\text{sin}^2(\phi)\Big)\text{d}\phi \nonumber\\
    &\qquad\qquad\qquad+ \frac{1}{4\pi A\omega_0^2}\oint\Big(D_{(A,\phi)}\text{cos}^2(\phi)\Big)\text{d}\phi\Bigg)\Bigg\rvert_{A=A_0}, \\
    &\sqrt{D_A} = \frac{1}{\sqrt{2\pi}\omega_0}\sqrt{\oint\left(\sqrt{D_{(A,\phi)}}\text{sin}(\phi)\right)^2\dd\phi}\Bigg\rvert_{A=A_0},
\end{align}
Substituting these into the Eq.~\ref{A_diff_approx}, one arrives at the much simpler equation
\begin{equation}
    \text{d}A = -\gamma_A\left(A-A_0\right)\text{d}t + \sqrt{D_A}\text{d}w_A.
\end{equation}
This equation describes the amplitude of the system as an Ornstein–Uhlenbeck process~\cite{Gardiner}.

We can perform a similar analysis on the equation of motion for the phase:
\begin{align}
    \text{d}\phi =& -\omega_0\text{d}t -\gamma_x\text{cos}(\phi)\text{sin}(\phi)\text{d}t +\frac{\text{cos}\phi F}{\omega_0A}\text{d}t \nonumber\\
    &+ \frac{\text{cos}(\phi)\sqrt{D_x}}{A\omega_0}\text{d}w - \frac{\text{cos}(\phi)\text{sin}(\phi)D_x}{2A^2\omega_0^2}\text{d}t.
\end{align}
We will assume that the oscillator's natural frequency is the dominate term in the deterministic part of the equation. The phase evolution is given by
\begin{equation}
    \label{eq:full phase}
    \text{d}\phi = -\omega_0\text{d}t + \frac{\text{cos}(\phi)\sqrt{D_x}}{A\omega_0}\text{d}w.
\end{equation}
Assuming sufficiently large $\omega_0^2A/\sqrt{D_x}$ we can linearise Eq.~\eqref{eq:full phase} by replacing the stochastic term prefactor with its RMS over a single cycle so that
\begin{equation}
    \text{d}\phi = -\omega_0\text{d}t + \sqrt{D_\phi}\text{d}w_\phi,
\end{equation}
here $\dd w_\phi$ is the effective Wiener increment originating from the linearisation. The diffusion term is given by
\begin{equation}
  \label{D_phi}
    \sqrt{D_\phi} = \frac{1}{A\omega_0\sqrt{2\pi}}\sqrt{\oint \left[\text{cos}^2(\phi)D_x\right]\text{d}\phi}\Bigg\rvert_{A=A_0}.
\end{equation}
Note that in the main text near Eqs.~\eqref{amplitude_eqn} and~\eqref{phase_eqn} we have replaced $\phi \to -\phi$ merely for clarity of presentation. 

\begin{figure}
	\begin{center}
		
		\includegraphics[width=\columnwidth,height=6.5cm]{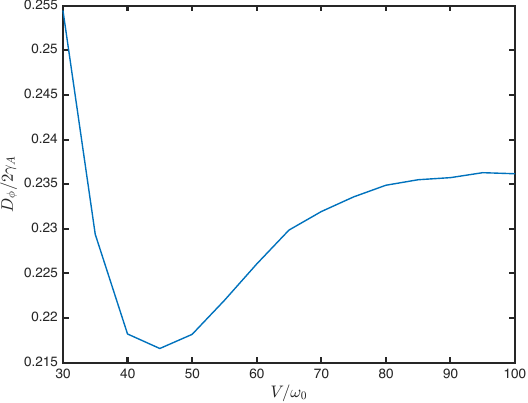}
		
		\caption{Ratio between the phase diffusion coefficient $D_\phi$ and the amplitude damping rate $\gamma_A$ computed for small fluctuations away from the ideal limit cycle. The relatively small values indicate that phase coherence on the limit cycle is long-lived compared to the rate of amplitude variation. Parameters are the same as Fig.~\ref{fig:Trajectory}. \label{fig:Ratio}}
		
	\end{center}
\end{figure} 

Since the amplitude and phase noise are independent within our approximations, it is straightforward to compute the steady-state position autocorrelation function~\cite{Gardiner}
\begin{align}
	\label{x_autocorrelation}
   \mathbb{E}[x(t)x(0)] -    \mathbb{E}[x(0)]^2  & = \mathbb{E}[A(t)A(0)]  \mathbb{E}[\cos \phi(t) \cos\phi(0)] \notag \\
&=   \left (A_0^2 + \frac{D_A \ee^{-2\gamma_A t}}{2\gamma_A}\right ) \frac{\cos(\omega_0 t) \ee^{-D_\phi t/2}}{2}. 
\end{align}
This function exhibits oscillatory decay at a rate that is ultimately dictated by the phase decoherence time, $D_\phi^{-1}$. This conclusion holds irrespectively of the scale, $D_A$, and regression rate, $\gamma_A$, of amplitude fluctuations. This point is discussed in more detail at the end of Appendix~\ref{app:corr_func}. 

To compare the characteristic timescales of amplitude and phase fluctuations that result from this description, we numerically compute the quantities $\gamma_A$ and $D_\phi$ from Eqs.~\eqref{gamma_A} and \eqref{D_phi}, respectively. Their ratio is plotted in Fig.~\ref{fig:Ratio} as a function of the voltage bias, for values of $V$ above and beyond the threshold for self-oscillations. We see that for all voltages considered, we have $D_\phi/4\gamma_A \ll 1$, implying that phase coherence lasts longer than the characteristic timescale of amplitude variations.

\section{\label{app:Bistable} Bistable model of large amplitude fluctuations}

In this appendix, we derive the autocorrelation function~\eqref{Bi_Corr} for a bistable current that switches between two states, finding that it reproduces the features seen in Fig.~\ref{fig:Two_Point_Corr} when the switching process is sufficiently slow. Concretely, consider a stochastic process $I(t)$ that switches randomly between two different oscillatory states labelled by $i=1,2$, giving rise to stochastic sub-signals $I_1(t)$ and $I_2(t)$. We are interested in the steady-state correlation function 
\begin{equation}
    C(t) = \mathbb{E}[I(t)I(0)] -\mathbb{E}[I(0)]^2.
\end{equation}
The average signal is 
\begin{equation}\label{bistable_average_signal}
    \mathbb{E}[I(0)] = p_1 \overline{I_1} + p_2 \overline{I_2},
\end{equation}
where $p_i$ is the steady-state probability to be in state $i$ and $\overline{I_i} = \mathbb{E}[I_i(0)]$ is the steady-state average of the corresponding sub-signal. Assuming that switching between the two states occurs independently of the value of $I_i(t)$, we can write
\begin{equation}
\label{tele_corr}
    \mathbb{E}\left[I(t)I(0)\right] = \sum_{i,j=1}^2p_{j|i}(t)p_i\mathbb{E}\left[I_j(t)I_i(0)\right],
\end{equation}
where $p_{j|i}(t)$ is the conditional probability to be in state $j$ at time $t$, given state $i$ at $t=0$. Here we take $t>0$ without loss of generality, since the stationary correlation function obeys $C(t)=C(-t)$. We further assume that the two sub-signals are statistically independent from each other, i.e.
\begin{equation}
    \label{subsignal_independence}
    \mathbb{E}[I_1(t)I_2(t')] =\mathbb{E}[I_1(t)]\mathbb{E}[I_2(t')].
\end{equation}
Now, splitting the sum in Eq.~\eqref{tele_corr} into diagonal ($i=j$) and off-diagonal ($i\neq j$) terms, and subtracting the square of Eq.~\eqref{bistable_average_signal}, we arrive at Eqs.~\eqref{Bi_Corr} and~\eqref{C_switch}, i.e.
\begin{align}
   &C(t) =\sum_{i=1}^2p_ip_{i|i}(t)C_i(t) + C_{1\leftrightarrow 2}(t), \label{Bi_Corr_app} \\ 
   & C_{1\leftrightarrow 2}(t) = \sum_{i,j=1}^2 p_i \left[p_{j|i}(t)-p_j\right]\overline{I_i}\, \overline{I_j}, \label{C_switch_app}
\end{align}
where $C_i(t) = \mathbb{E}[I_i(t)I_i(0)] -\overline{I_i}^2$ is the auto-correlation function of $I_i(t)$. 

The switching term $ C_{1\leftrightarrow 2}(t) $ has the properties expected of an autocorrelation function: in particular, it vanishes as $t\to \infty$ since $p_{j|i}(t) \to p_j$ at long times. Using the conservation of probability, $\sum_i p_i = \sum_j p_{j|i}(t) = 1$, one can also show that 
\begin{equation}
\label{C_switch_explicit}
  C_{1\leftrightarrow 2}(t) = \left (\overline{I_2} - \overline{I_1}\right )\left (p_1\overline{I}_1\left [p_{2|1}(t)-p_2\right ] - p_2\overline{I_2}\left [p_{1|2}(t)-p_1\right ]\right ).
\end{equation}
Therefore, the switching term is non-zero only if $\overline{I_2} \neq \overline{I_1}$, i.e.~the two sub-signals have a different mean value. If this is the case, and if the switching probabilities $p_{j|i}(t)$ evolve on a slow timescale compared to $C_i(t)$, then Eq.~\eqref{Bi_Corr_app} will exhibit two well-separated timescales as in Fig.~\ref{fig:Two_Point_Corr}.

To demonstrate this explicitly, we assume that the switching occurs via a simple telegraph process described by the master equation
\begin{equation}
\label{telegraph_process}
 \dot{p}_1(t) = \lambda_2p_2(t) -\lambda_1p_1(t)  = - \dot{p}_2(t),
\end{equation}
where $\lambda_i$ are transition rates between the sub-signals. The steady-state probabilities are given by
\begin{equation}
        \begin{pmatrix}
 p_1\\p_2
    \end{pmatrix} =\frac{1}{\lambda_1+\lambda_2}
        \begin{pmatrix}
        \lambda_2\\\lambda_1\label{Tele_Mean}
    \end{pmatrix},
\end{equation}
The transition probability of switching from state $i$ to state $j\neq i$ after a time $t>0$ is given by the solution of Eq.~\eqref{telegraph_process} with $p_i(0) = 1$, i.e.
\begin{equation}
\label{transition_prob}
	 p_{j|i}(t) = \frac{\lambda_i}{\lambda_1+\lambda_2}\left (1 - \ee^{-(\lambda_1+\lambda_2)t} \right ).
\end{equation}
Substituting these results into Eq.~\eqref{C_switch_explicit} gives the two-point correlation function of the telegraph process
\begin{equation}
    \label{telegraph_corr}
C_{1\leftrightarrow 2}(t) = \frac{\left (\overline{I_2} - \overline{I_1}\right )^2 \lambda_1\lambda_2 \ee^{-(\lambda_1+\lambda_2)|t|}}{\left (\lambda_1+\lambda_2\right )^2}.
\end{equation}
Therefore, the total correlation function [Eq.~\eqref{Bi_Corr_app}] is the sum of two terms: an oscillatory function determined by the $C_i(t)$, which decays due to amplitude and phase fluctuations within each state, and a pure exponential decay given by Eq.~\eqref{telegraph_corr}. So long as the total switching rate $\lambda_1 + \lambda_2$ is slow compared to the decay of the $C_i(t)$, the latter contribution will survive after the oscillations have died out. This is precisely the qualitative behaviour above threshold seen in Fig.~\ref{fig:Two_Point_Corr}. 

This model also predicts a large zero-frequency component to the current noise, consistent with our observations in Sec.~\ref{sec:power_spectrum}. From Eq.~\eqref{power_spectrum} one has~\cite{landi_current_2023}
\begin{equation}
	\mathfrak{S}(0) = 2\int_0^\infty \dd t\, C(t).
\end{equation}
Therefore the switching term contributes a term that scales as 
\begin{equation}
	\label{S_0_scaling}
		\mathfrak{S}(0)  \sim \frac{\lambda_1\lambda_2}{(\lambda_1+\lambda_2)^3},
\end{equation}
which diverges as $\lambda_i \to 0$. For example, if $\lambda_1=\lambda_2 = \lambda$ then we simply have  $	\mathfrak{S}(0) \sim \lambda^{-1}$, where $\lambda^{-1}$ is the lifetime of each state. 

Note that other models of slow amplitude fluctuations give rise to similar behaviour. However, it is crucial that the  cycle-averaged signal depends on the amplitude, which is reflected here by the condition that $\overline{I_1} \neq \overline{I_2}$. For example, the model of independent amplitude and phase fluctuations developed in Appendix~\ref{app:OU_derivation} yields a position autocorrelation function given by Eq.~\eqref{x_autocorrelation}. This does not show a separation of timescales, even when the amplitude fluctuations are much slower than phase fluctuations, i.e.~$D_A, \gamma_A \ll D_\phi$. This is because the average of $x(t) = A(t)\cos\phi (t)$ around one cycle is zero, independently of the value of $A(t)$. 

To illustrate the importance of an amplitude-dependent cycle average, let us instead consider a model of stochastic oscillations with an offset:
\begin{equation}
    \label{offset_model}
 	I(t) = A(t)\left \{\cos[\phi(t)] - b \right \} +c,
\end{equation}
where $b$ and $c$ are constants, and $A(t)$ and $\phi(t)$ are described by an Ornstein-Uhlenbeck process [Eq.~\eqref{amplitude_eqn}] and a biased diffusion process [Eq.~\eqref{phase_eqn}], respectively. The resulting stochastic dynamics is qualitatively similar to the behaviour seen in Fig.~\ref{fig:Trajectory} in regions where the peak-to-peak amplitude is large, with $c$ representing the maximum current and the scale factor $b$ determining how much oscillations reduce the current on average. In particular, the cycle average of this signal is now amplitude dependent and given by $\mathbb{E}_\phi[I(t)] = c- bA(t)$. We find the autocorrelation function for this model is given by 
\begin{equation}
	\label{amplitude_fluct_correlation}
	 C(t) = \left (A_0^2 + \frac{D_A \ee^{-2\gamma_A t}}{2\gamma_A}\right ) \frac{\cos(\omega_0 t) \ee^{-D_\phi t/2}}{2} + \frac{b^2 D_A  \ee^{-2\gamma_A t}}{2\gamma_A}.
\end{equation}
The first term above is identical to Eq.~\eqref{x_autocorrelation}, but now another term appears that depends only on the amplitude fluctuations. This latter term can be understood as a continuous analogue of Eq.~\eqref{C_switch_app}, where now $b$ instead of $\overline{I_1}- \overline{I_2}$ determines how amplitude fluctuations affect the cycle-averaged signal.

\section{Allan variance for a renewal clock\label{app:allan}}

In this appendix we derive the asymptotic Allan variance for a clock whose ticks are generated by a renewal process. Our derivation takes a different approach from that of Ref.~\cite{silva_ticking_2023}, although the results are equivalent up to a rescaling. Let $n(t)$ denote the number of ticks generated after a coordinate time $t$ has elapsed. Let us assume that $n(t)$ is a renewal process for which the mean and variance are well known~\cite{Cox_1967}
\begin{align}
&	\mathbb{E}\left [n(t)\right ] = \frac{t}{\mu} + \mathcal{O}(t^0), \label{renewal_mean}\\
\label{renewal_var}
&	\mathbb{E}\left [n(t)^2 \right ] - \mathbb{E}[n(t)]^2 = \frac{\sigma^2 t}{\mu^3}+ \mathcal{O}(t^0),
 \end{align}
 where $\mu$ and $\sigma$ are the mean and variance of the waiting-time distribution. 
 
 The clock reading at time $t$ is taken to be
\begin{equation}
	\label{renewal_clock_reading}
	\theta(t) = \mu n(t),
\end{equation}
which is the  ``asymptotically unbiased'' choice since then $\mathbb{E}[\theta(t)] = t$ at long times. The fractional frequency deviation averaged over a time $T$ is then, from Eq.~\eqref{av_frac_freq_diff},
\begin{equation}
	\label{renewal_av_frac_diff}
	\bar{y}(t,T) = \frac{\mu}{T}\left [ n(t+T) - n(t)\right ].
\end{equation}
Now, the Allan variance is defined as
\begin{align}
		\sigma^2_y(T) & = \frac{1}{2}\mathbb{E}\left\{\left[\bar{y}(t+T,T) - \bar{y}(t,T)\right]^2 \right\}  \notag \\
		& = \frac{1}{2}\mathbb{E}\left[\bar{y}(t+T,T)^2\right ] +\frac{1}{2}\mathbb{E}\left[\bar{y}(t,T)^2\right ]\notag \\ & \quad - \mathbb{E}\left[\bar{y}(t+T,T)\bar{y}(t,T) \right] \notag \\
		& \underset{T\to \infty}{\longrightarrow} \mathbb{E}\left[\bar{y}(t,T)^2\right ] - \mathbb{E}\left[\bar{y}(t,T) \right]^2
\end{align}
On the second line, the first two terms are equal because the process is stationary. The third line then follows in the limit of large $T$ because $\bar{y}(t+T,T)$ and $\bar{y}(t,T)$ become statistically independent, i.e.~the number of renewals in time interval $[t,t+T]$ becomes independent of the number of renewals in the following time interval $[t+T,t+2T]$. Therefore, the Allan variance for large $T$ is (up to scale factors) nothing but the variance of, $n(t+T)- n(t)$, i.e.~the number of renewals within a time interval of duration $T$. Analogously to Eq.~\eqref{renewal_var}, this result is known asymptotically to be ${\rm Var}\left [ n(t+T) - n(t)\right ] = \sigma^2 T/\mu^3 + \mathcal{O}(T^0)$~\cite{Cox_1967}. Finally, this yields the asymptotic Allan variance for a renewal process
\begin{equation}
	\label{Allan_asymptotic}
		\sigma^2_y(T) \longrightarrow \frac{\sigma^2}{\mu T},
\end{equation}
which is equivalent to Eq.~\eqref{renewal_allan_variance}.

\bibliographystyle{apsrev4-2}
\bibliography{references_11.bib,bib_7.bib}

\begin{thebibliography}{76}%
\makeatletter
\providecommand \@ifxundefined [1]{%
 \@ifx{#1\undefined}
}%
\providecommand \@ifnum [1]{%
 \ifnum #1\expandafter \@firstoftwo
 \else \expandafter \@secondoftwo
 \fi
}%
\providecommand \@ifx [1]{%
 \ifx #1\expandafter \@firstoftwo
 \else \expandafter \@secondoftwo
 \fi
}%
\providecommand \natexlab [1]{#1}%
\providecommand \enquote  [1]{``#1''}%
\providecommand \bibnamefont  [1]{#1}%
\providecommand \bibfnamefont [1]{#1}%
\providecommand \citenamefont [1]{#1}%
\providecommand \href@noop [0]{\@secondoftwo}%
\providecommand \href [0]{\begingroup \@sanitize@url \@href}%
\providecommand \@href[1]{\@@startlink{#1}\@@href}%
\providecommand \@@href[1]{\endgroup#1\@@endlink}%
\providecommand \@sanitize@url [0]{\catcode `\\12\catcode `\$12\catcode
  `\&12\catcode `\#12\catcode `\^12\catcode `\_12\catcode `\%12\relax}%
\providecommand \@@startlink[1]{}%
\providecommand \@@endlink[0]{}%
\providecommand \url  [0]{\begingroup\@sanitize@url \@url }%
\providecommand \@url [1]{\endgroup\@href {#1}{\urlprefix }}%
\providecommand \urlprefix  [0]{URL }%
\providecommand \Eprint [0]{\href }%
\providecommand \doibase [0]{https://doi.org/}%
\providecommand \selectlanguage [0]{\@gobble}%
\providecommand \bibinfo  [0]{\@secondoftwo}%
\providecommand \bibfield  [0]{\@secondoftwo}%
\providecommand \translation [1]{[#1]}%
\providecommand \BibitemOpen [0]{}%
\providecommand \bibitemStop [0]{}%
\providecommand \bibitemNoStop [0]{.\EOS\space}%
\providecommand \EOS [0]{\spacefactor3000\relax}%
\providecommand \BibitemShut  [1]{\csname bibitem#1\endcsname}%
\let\auto@bib@innerbib\@empty
\bibitem [{\citenamefont {Stray}\ \emph {et~al.}(2022)\citenamefont {Stray},
  \citenamefont {Lamb}, \citenamefont {Kaushik}, \citenamefont {Vovrosh},
  \citenamefont {Rodgers}, \citenamefont {Winch}, \citenamefont {Hayati},
  \citenamefont {Boddice}, \citenamefont {Stabrawa}, \citenamefont {Niggebaum},
  \citenamefont {Langlois}, \citenamefont {Lien}, \citenamefont {Lellouch},
  \citenamefont {Roshanmanesh}, \citenamefont {Ridley}, \citenamefont
  {de~Villiers}, \citenamefont {Brown}, \citenamefont {Cross}, \citenamefont
  {Tuckwell}, \citenamefont {Faramarzi}, \citenamefont {Metje}, \citenamefont
  {Bongs},\ and\ \citenamefont {Holynski}}]{stray_quantum_2022}%
  \BibitemOpen
  \bibfield  {author} {\bibinfo {author} {\bibfnamefont {B.}~\bibnamefont
  {Stray}}, \bibinfo {author} {\bibfnamefont {A.}~\bibnamefont {Lamb}},
  \bibinfo {author} {\bibfnamefont {A.}~\bibnamefont {Kaushik}}, \bibinfo
  {author} {\bibfnamefont {J.}~\bibnamefont {Vovrosh}}, \bibinfo {author}
  {\bibfnamefont {A.}~\bibnamefont {Rodgers}}, \bibinfo {author} {\bibfnamefont
  {J.}~\bibnamefont {Winch}}, \bibinfo {author} {\bibfnamefont
  {F.}~\bibnamefont {Hayati}}, \bibinfo {author} {\bibfnamefont
  {D.}~\bibnamefont {Boddice}}, \bibinfo {author} {\bibfnamefont
  {A.}~\bibnamefont {Stabrawa}}, \bibinfo {author} {\bibfnamefont
  {A.}~\bibnamefont {Niggebaum}}, \bibinfo {author} {\bibfnamefont
  {M.}~\bibnamefont {Langlois}}, \bibinfo {author} {\bibfnamefont {Y.-H.}\
  \bibnamefont {Lien}}, \bibinfo {author} {\bibfnamefont {S.}~\bibnamefont
  {Lellouch}}, \bibinfo {author} {\bibfnamefont {S.}~\bibnamefont
  {Roshanmanesh}}, \bibinfo {author} {\bibfnamefont {K.}~\bibnamefont
  {Ridley}}, \bibinfo {author} {\bibfnamefont {G.}~\bibnamefont {de~Villiers}},
  \bibinfo {author} {\bibfnamefont {G.}~\bibnamefont {Brown}}, \bibinfo
  {author} {\bibfnamefont {T.}~\bibnamefont {Cross}}, \bibinfo {author}
  {\bibfnamefont {G.}~\bibnamefont {Tuckwell}}, \bibinfo {author}
  {\bibfnamefont {A.}~\bibnamefont {Faramarzi}}, \bibinfo {author}
  {\bibfnamefont {N.}~\bibnamefont {Metje}}, \bibinfo {author} {\bibfnamefont
  {K.}~\bibnamefont {Bongs}},\ and\ \bibinfo {author} {\bibfnamefont
  {M.}~\bibnamefont {Holynski}},\ }\href
  {https://www.nature.com/articles/s41586-021-04315-3} {\bibfield  {journal}
  {\bibinfo  {journal} {Nature}\ }\textbf {\bibinfo {volume} {602}},\ \bibinfo
  {pages} {590} (\bibinfo {year} {2022})}\BibitemShut {NoStop}%
\bibitem [{\citenamefont {Bothwell}\ \emph {et~al.}(2022)\citenamefont
  {Bothwell}, \citenamefont {Kennedy}, \citenamefont {Aeppli}, \citenamefont
  {Kedar}, \citenamefont {Robinson}, \citenamefont {Oelker}, \citenamefont
  {Staron},\ and\ \citenamefont {Ye}}]{bothwell_resolving_2022}%
  \BibitemOpen
  \bibfield  {author} {\bibinfo {author} {\bibfnamefont {T.}~\bibnamefont
  {Bothwell}}, \bibinfo {author} {\bibfnamefont {C.~J.}\ \bibnamefont
  {Kennedy}}, \bibinfo {author} {\bibfnamefont {A.}~\bibnamefont {Aeppli}},
  \bibinfo {author} {\bibfnamefont {D.}~\bibnamefont {Kedar}}, \bibinfo
  {author} {\bibfnamefont {J.~M.}\ \bibnamefont {Robinson}}, \bibinfo {author}
  {\bibfnamefont {E.}~\bibnamefont {Oelker}}, \bibinfo {author} {\bibfnamefont
  {A.}~\bibnamefont {Staron}},\ and\ \bibinfo {author} {\bibfnamefont
  {J.}~\bibnamefont {Ye}},\ }\href
  {https://www.nature.com/articles/s41586-021-04349-7} {\bibfield  {journal}
  {\bibinfo  {journal} {Nature}\ }\textbf {\bibinfo {volume} {602}},\ \bibinfo
  {pages} {420} (\bibinfo {year} {2022})},\ \Eprint
  {https://arxiv.org/abs/2109.12238} {arXiv:2109.12238} \BibitemShut {NoStop}%
\bibitem [{\citenamefont {Salecker}\ and\ \citenamefont
  {Wigner}(1958)}]{salecker_quantum_1958}%
  \BibitemOpen
  \bibfield  {author} {\bibinfo {author} {\bibfnamefont {H.}~\bibnamefont
  {Salecker}}\ and\ \bibinfo {author} {\bibfnamefont {E.~P.}\ \bibnamefont
  {Wigner}},\ }\href {https://link.aps.org/doi/10.1103/PhysRev.109.571}
  {\bibfield  {journal} {\bibinfo  {journal} {Phys. Rev.}\ }\textbf {\bibinfo
  {volume} {109}},\ \bibinfo {pages} {571} (\bibinfo {year}
  {1958})}\BibitemShut {NoStop}%
\bibitem [{\citenamefont {Peres}(1980)}]{peres_measurement_1980}%
  \BibitemOpen
  \bibfield  {author} {\bibinfo {author} {\bibfnamefont {A.}~\bibnamefont
  {Peres}},\ }\href {https://doi.org/10.1119/1.12061} {\bibfield  {journal}
  {\bibinfo  {journal} {Am. J. Phys.}\ }\textbf {\bibinfo {volume} {48}},\
  \bibinfo {pages} {552} (\bibinfo {year} {1980})}\BibitemShut {NoStop}%
\bibitem [{\citenamefont {Page}\ and\ \citenamefont
  {Wootters}(1983)}]{page_evolution_1983}%
  \BibitemOpen
  \bibfield  {author} {\bibinfo {author} {\bibfnamefont {D.~N.}\ \bibnamefont
  {Page}}\ and\ \bibinfo {author} {\bibfnamefont {W.~K.}\ \bibnamefont
  {Wootters}},\ }\href {https://link.aps.org/doi/10.1103/PhysRevD.27.2885}
  {\bibfield  {journal} {\bibinfo  {journal} {Phys. Rev. D}\ }\textbf {\bibinfo
  {volume} {27}},\ \bibinfo {pages} {2885} (\bibinfo {year}
  {1983})}\BibitemShut {NoStop}%
\bibitem [{\citenamefont {Janzing}\ and\ \citenamefont
  {Beth}(2003)}]{janzing_quasi-order_2003}%
  \BibitemOpen
  \bibfield  {author} {\bibinfo {author} {\bibfnamefont {D.}~\bibnamefont
  {Janzing}}\ and\ \bibinfo {author} {\bibfnamefont {T.}~\bibnamefont {Beth}},\
  }\href {https://doi.org/10.1109/TIT.2002.806162} {\bibfield  {journal}
  {\bibinfo  {journal} {IEEE Transactions on Information Theory}\ }\textbf
  {\bibinfo {volume} {49}},\ \bibinfo {pages} {230} (\bibinfo {year} {2003})},\
  \Eprint {https://arxiv.org/abs/0112138} {arXiv:0112138} \BibitemShut
  {NoStop}%
\bibitem [{\citenamefont {Rankovi{\'c}}\ \emph {et~al.}(2015)\citenamefont
  {Rankovi{\'c}}, \citenamefont {Liang},\ and\ \citenamefont
  {Renner}}]{rankovic_quantum_2015}%
  \BibitemOpen
  \bibfield  {author} {\bibinfo {author} {\bibfnamefont {S.}~\bibnamefont
  {Rankovi{\'c}}}, \bibinfo {author} {\bibfnamefont {Y.-C.}\ \bibnamefont
  {Liang}},\ and\ \bibinfo {author} {\bibfnamefont {R.}~\bibnamefont
  {Renner}},\ }\href {http://arxiv.org/abs/1506.01373} {\bibinfo {title}
  {Quantum clocks and their synchronisation - the {Alternate} {Ticks} {Game}}}
  (\bibinfo {year} {2015}),\ \bibinfo {note} {arXiv:1506.01373}\BibitemShut
  {NoStop}%
\bibitem [{\citenamefont {Woods}\ \emph {et~al.}(2019)\citenamefont {Woods},
  \citenamefont {Silva},\ and\ \citenamefont
  {Oppenheim}}]{woods_autonomous_2019}%
  \BibitemOpen
  \bibfield  {author} {\bibinfo {author} {\bibfnamefont {M.~P.}\ \bibnamefont
  {Woods}}, \bibinfo {author} {\bibfnamefont {R.}~\bibnamefont {Silva}},\ and\
  \bibinfo {author} {\bibfnamefont {J.}~\bibnamefont {Oppenheim}},\ }\href
  {https://doi.org/10.1007/s00023-018-0736-9} {\bibfield  {journal} {\bibinfo
  {journal} {Annales Henri Poincar{\'e}}\ }\textbf {\bibinfo {volume} {20}},\
  \bibinfo {pages} {125} (\bibinfo {year} {2019})},\ \Eprint
  {https://arxiv.org/abs/1607.04591} {arXiv:1607.04591} \BibitemShut {NoStop}%
\bibitem [{\citenamefont {Woods}(2021)}]{woods_autonomous_2021}%
  \BibitemOpen
  \bibfield  {author} {\bibinfo {author} {\bibfnamefont {M.~P.}\ \bibnamefont
  {Woods}},\ }\href {https://quantum-journal.org/papers/q-2021-01-17-381/}
  {\bibfield  {journal} {\bibinfo  {journal} {Quantum}\ }\textbf {\bibinfo
  {volume} {5}},\ \bibinfo {pages} {381} (\bibinfo {year} {2021})},\ \Eprint
  {https://arxiv.org/abs/2005.04628} {arXiv:2005.04628} \BibitemShut {NoStop}%
\bibitem [{\citenamefont {Woods}\ \emph {et~al.}(2022)\citenamefont {Woods},
  \citenamefont {Silva}, \citenamefont {P{\"u}tz}, \citenamefont {Stupar},\
  and\ \citenamefont {Renner}}]{woods_quantum_2022}%
  \BibitemOpen
  \bibfield  {author} {\bibinfo {author} {\bibfnamefont {M.~P.}\ \bibnamefont
  {Woods}}, \bibinfo {author} {\bibfnamefont {R.}~\bibnamefont {Silva}},
  \bibinfo {author} {\bibfnamefont {G.}~\bibnamefont {P{\"u}tz}}, \bibinfo
  {author} {\bibfnamefont {S.}~\bibnamefont {Stupar}},\ and\ \bibinfo {author}
  {\bibfnamefont {R.}~\bibnamefont {Renner}},\ }\href
  {https://link.aps.org/doi/10.1103/PRXQuantum.3.010319} {\bibfield  {journal}
  {\bibinfo  {journal} {PRX Quantum}\ }\textbf {\bibinfo {volume} {3}},\
  \bibinfo {pages} {010319} (\bibinfo {year} {2022})},\ \Eprint
  {https://arxiv.org/abs/1806.00491} {arXiv:1806.00491} \BibitemShut {NoStop}%
\bibitem [{\citenamefont {Erker}\ \emph {et~al.}(2017)\citenamefont {Erker},
  \citenamefont {Mitchison}, \citenamefont {Silva}, \citenamefont {Woods},
  \citenamefont {Brunner},\ and\ \citenamefont {Huber}}]{Erker_2017}%
  \BibitemOpen
  \bibfield  {author} {\bibinfo {author} {\bibfnamefont {P.}~\bibnamefont
  {Erker}}, \bibinfo {author} {\bibfnamefont {M.~T.}\ \bibnamefont
  {Mitchison}}, \bibinfo {author} {\bibfnamefont {R.}~\bibnamefont {Silva}},
  \bibinfo {author} {\bibfnamefont {M.~P.}\ \bibnamefont {Woods}}, \bibinfo
  {author} {\bibfnamefont {N.}~\bibnamefont {Brunner}},\ and\ \bibinfo {author}
  {\bibfnamefont {M.}~\bibnamefont {Huber}},\ }\href
  {https://doi.org/10.1103/PhysRevX.7.031022} {\bibfield  {journal} {\bibinfo
  {journal} {Phys. Rev. X}\ }\textbf {\bibinfo {volume} {7}},\ \bibinfo {pages}
  {031022} (\bibinfo {year} {2017})}\BibitemShut {NoStop}%
\bibitem [{\citenamefont {Barato}\ and\ \citenamefont
  {Seifert}(2016)}]{Barato_2016}%
  \BibitemOpen
  \bibfield  {author} {\bibinfo {author} {\bibfnamefont {A.~C.}\ \bibnamefont
  {Barato}}\ and\ \bibinfo {author} {\bibfnamefont {U.}~\bibnamefont
  {Seifert}},\ }\href {https://doi.org/10.1103/PhysRevX.6.041053} {\bibfield
  {journal} {\bibinfo  {journal} {Phys. Rev. X}\ }\textbf {\bibinfo {volume}
  {6}},\ \bibinfo {pages} {041053} (\bibinfo {year} {2016})}\BibitemShut
  {NoStop}%
\bibitem [{\citenamefont {Milburn}(2020)}]{Milburn2020}%
  \BibitemOpen
  \bibfield  {author} {\bibinfo {author} {\bibfnamefont {G.~J.}\ \bibnamefont
  {Milburn}},\ }\href {https://doi.org/10.1080/00107514.2020.1837471}
  {\bibfield  {journal} {\bibinfo  {journal} {Contemp. Phys.}\ }\textbf
  {\bibinfo {volume} {61}},\ \bibinfo {pages} {69} (\bibinfo {year} {2020})},\
  \Eprint {https://arxiv.org/abs/2007.02217} {arXiv:2007.02217} \BibitemShut
  {NoStop}%
\bibitem [{\citenamefont {Schwarzhans}\ \emph {et~al.}(2021)\citenamefont
  {Schwarzhans}, \citenamefont {Lock}, \citenamefont {Erker}, \citenamefont
  {Friis},\ and\ \citenamefont {Huber}}]{Schwarzhans2021}%
  \BibitemOpen
  \bibfield  {author} {\bibinfo {author} {\bibfnamefont {E.}~\bibnamefont
  {Schwarzhans}}, \bibinfo {author} {\bibfnamefont {M.~P.~E.}\ \bibnamefont
  {Lock}}, \bibinfo {author} {\bibfnamefont {P.}~\bibnamefont {Erker}},
  \bibinfo {author} {\bibfnamefont {N.}~\bibnamefont {Friis}},\ and\ \bibinfo
  {author} {\bibfnamefont {M.}~\bibnamefont {Huber}},\ }\href
  {https://doi.org/10.1103/PhysRevX.11.011046} {\bibfield  {journal} {\bibinfo
  {journal} {Phys. Rev. X}\ }\textbf {\bibinfo {volume} {11}},\ \bibinfo
  {pages} {011046} (\bibinfo {year} {2021})},\ \Eprint
  {https://arxiv.org/abs/2007.01307} {arXiv:2007.01307} \BibitemShut {NoStop}%
\bibitem [{\citenamefont {Dost}\ and\ \citenamefont
  {Woods}(2023)}]{dost_quantum_2023}%
  \BibitemOpen
  \bibfield  {author} {\bibinfo {author} {\bibfnamefont {A.~P.~T.}\
  \bibnamefont {Dost}}\ and\ \bibinfo {author} {\bibfnamefont {M.~P.}\
  \bibnamefont {Woods}},\ }\href {http://arxiv.org/abs/2303.10029} {\bibinfo
  {title} {Quantum advantages in timekeeping: dimensional advantage, entropic
  advantage and how to realise them via {Berry} phases and ultra-regular
  spontaneous emission}} (\bibinfo {year} {2023}),\ \bibinfo {note}
  {arXiv:2303.10029}\BibitemShut {NoStop}%
\bibitem [{\citenamefont {Meier}\ \emph {et~al.}(2023)\citenamefont {Meier},
  \citenamefont {Schwarzhans}, \citenamefont {Erker},\ and\ \citenamefont
  {Huber}}]{meier_fundamental_2023}%
  \BibitemOpen
  \bibfield  {author} {\bibinfo {author} {\bibfnamefont {F.}~\bibnamefont
  {Meier}}, \bibinfo {author} {\bibfnamefont {E.}~\bibnamefont {Schwarzhans}},
  \bibinfo {author} {\bibfnamefont {P.}~\bibnamefont {Erker}},\ and\ \bibinfo
  {author} {\bibfnamefont {M.}~\bibnamefont {Huber}},\ }\href
  {https://doi.org/10.1103/PhysRevLett.131.220201} {\bibfield  {journal}
  {\bibinfo  {journal} {Phys. Rev. Lett.}\ }\textbf {\bibinfo {volume} {131}},\
  \bibinfo {pages} {220201} (\bibinfo {year} {2023})}\BibitemShut {NoStop}%
\bibitem [{\citenamefont {Barato}\ and\ \citenamefont
  {Seifert}(2015)}]{barato_thermodynamic_2015}%
  \BibitemOpen
  \bibfield  {author} {\bibinfo {author} {\bibfnamefont {A.~C.}\ \bibnamefont
  {Barato}}\ and\ \bibinfo {author} {\bibfnamefont {U.}~\bibnamefont
  {Seifert}},\ }\href {https://link.aps.org/doi/10.1103/PhysRevLett.114.158101}
  {\bibfield  {journal} {\bibinfo  {journal} {Phys. Rev. Letters}\ }\textbf
  {\bibinfo {volume} {114}},\ \bibinfo {pages} {158101} (\bibinfo {year}
  {2015})},\ \Eprint {https://arxiv.org/abs/1502.05944} {arXiv:1502.05944}
  \BibitemShut {NoStop}%
\bibitem [{\citenamefont {Gingrich}\ \emph {et~al.}(2016)\citenamefont
  {Gingrich}, \citenamefont {Horowitz}, \citenamefont {Perunov},\ and\
  \citenamefont {England}}]{gingrich_dissipation_2016}%
  \BibitemOpen
  \bibfield  {author} {\bibinfo {author} {\bibfnamefont {T.~R.}\ \bibnamefont
  {Gingrich}}, \bibinfo {author} {\bibfnamefont {J.~M.}\ \bibnamefont
  {Horowitz}}, \bibinfo {author} {\bibfnamefont {N.}~\bibnamefont {Perunov}},\
  and\ \bibinfo {author} {\bibfnamefont {J.~L.}\ \bibnamefont {England}},\
  }\href {https://link.aps.org/doi/10.1103/PhysRevLett.116.120601} {\bibfield
  {journal} {\bibinfo  {journal} {Phys. Rev. Letters}\ }\textbf {\bibinfo
  {volume} {116}},\ \bibinfo {pages} {120601} (\bibinfo {year} {2016})},\
  \Eprint {https://arxiv.org/abs/1512.02212} {arXiv:1512.02212} \BibitemShut
  {NoStop}%
\bibitem [{\citenamefont {Garrahan}(2017)}]{garrahan_simple_2017}%
  \BibitemOpen
  \bibfield  {author} {\bibinfo {author} {\bibfnamefont {J.~P.}\ \bibnamefont
  {Garrahan}},\ }\href {https://link.aps.org/doi/10.1103/PhysRevE.95.032134}
  {\bibfield  {journal} {\bibinfo  {journal} {Phys. Rev. E}\ }\textbf {\bibinfo
  {volume} {95}},\ \bibinfo {pages} {032134} (\bibinfo {year} {2017})},\
  \Eprint {https://arxiv.org/abs/1701.00539} {arXiv:1701.00539} \BibitemShut
  {NoStop}%
\bibitem [{\citenamefont {Terlizzi}\ and\ \citenamefont
  {Baiesi}(2018)}]{terlizzi_kinetic_2018}%
  \BibitemOpen
  \bibfield  {author} {\bibinfo {author} {\bibfnamefont {I.~D.}\ \bibnamefont
  {Terlizzi}}\ and\ \bibinfo {author} {\bibfnamefont {M.}~\bibnamefont
  {Baiesi}},\ }\href {https://dx.doi.org/10.1088/1751-8121/aaee34} {\bibfield
  {journal} {\bibinfo  {journal} {J. Phys. A: Mathematical and Theoretical}\
  }\textbf {\bibinfo {volume} {52}},\ \bibinfo {pages} {02LT03} (\bibinfo
  {year} {2018})}\BibitemShut {NoStop}%
\bibitem [{\citenamefont
  {Ptaszy{\'n}ski}(2018)}]{ptaszynski_coherence-enhanced_2018}%
  \BibitemOpen
  \bibfield  {author} {\bibinfo {author} {\bibfnamefont {K.}~\bibnamefont
  {Ptaszy{\'n}ski}},\ }\href
  {https://link.aps.org/doi/10.1103/PhysRevB.98.085425} {\bibfield  {journal}
  {\bibinfo  {journal} {Phys. Rev. B}\ }\textbf {\bibinfo {volume} {98}},\
  \bibinfo {pages} {085425} (\bibinfo {year} {2018})},\ \Eprint
  {https://arxiv.org/abs/1805.11301} {arXiv:1805.11301} \BibitemShut {NoStop}%
\bibitem [{\citenamefont {Agarwalla}\ and\ \citenamefont
  {Segal}(2018)}]{agarwalla_assessing_2018}%
  \BibitemOpen
  \bibfield  {author} {\bibinfo {author} {\bibfnamefont {B.~K.}\ \bibnamefont
  {Agarwalla}}\ and\ \bibinfo {author} {\bibfnamefont {D.}~\bibnamefont
  {Segal}},\ }\href {https://doi.org/10.1103/PhysRevB.98.155438} {\bibfield
  {journal} {\bibinfo  {journal} {Phys. Rev. B}\ }\textbf {\bibinfo {volume}
  {98}},\ \bibinfo {pages} {155438} (\bibinfo {year} {2018})},\ \Eprint
  {https://arxiv.org/abs/1806.05588} {arXiv:1806.05588} \BibitemShut {NoStop}%
\bibitem [{\citenamefont {Guarnieri}\ \emph {et~al.}(2019)\citenamefont
  {Guarnieri}, \citenamefont {Landi}, \citenamefont {Clark},\ and\
  \citenamefont {Goold}}]{guarnieri_thermodynamics_2019}%
  \BibitemOpen
  \bibfield  {author} {\bibinfo {author} {\bibfnamefont {G.}~\bibnamefont
  {Guarnieri}}, \bibinfo {author} {\bibfnamefont {G.~T.}\ \bibnamefont
  {Landi}}, \bibinfo {author} {\bibfnamefont {S.~R.}\ \bibnamefont {Clark}},\
  and\ \bibinfo {author} {\bibfnamefont {J.}~\bibnamefont {Goold}},\ }\href
  {https://link.aps.org/doi/10.1103/PhysRevResearch.1.033021} {\bibfield
  {journal} {\bibinfo  {journal} {Phys. Rev. Research}\ }\textbf {\bibinfo
  {volume} {1}},\ \bibinfo {pages} {033021} (\bibinfo {year} {2019})},\ \Eprint
  {https://arxiv.org/abs/1901.10428} {arXiv:1901.10428} \BibitemShut {NoStop}%
\bibitem [{\citenamefont {Malabarba}\ \emph {et~al.}(2015)\citenamefont
  {Malabarba}, \citenamefont {Short},\ and\ \citenamefont
  {Kammerlander}}]{malabarba_clock-driven_2015}%
  \BibitemOpen
  \bibfield  {author} {\bibinfo {author} {\bibfnamefont {A.~S.~L.}\
  \bibnamefont {Malabarba}}, \bibinfo {author} {\bibfnamefont {A.~J.}\
  \bibnamefont {Short}},\ and\ \bibinfo {author} {\bibfnamefont
  {P.}~\bibnamefont {Kammerlander}},\ }\href
  {https://dx.doi.org/10.1088/1367-2630/17/4/045027} {\bibfield  {journal}
  {\bibinfo  {journal} {NJP}\ }\textbf {\bibinfo {volume} {17}},\ \bibinfo
  {pages} {045027} (\bibinfo {year} {2015})},\ \Eprint
  {https://arxiv.org/abs/1412.1338} {arXiv:1412.1338} \BibitemShut {NoStop}%
\bibitem [{\citenamefont {Frenzel}\ \emph {et~al.}(2016)\citenamefont
  {Frenzel}, \citenamefont {Jennings},\ and\ \citenamefont
  {Rudolph}}]{frenzel_quasi-autonomous_2016}%
  \BibitemOpen
  \bibfield  {author} {\bibinfo {author} {\bibfnamefont {M.~F.}\ \bibnamefont
  {Frenzel}}, \bibinfo {author} {\bibfnamefont {D.}~\bibnamefont {Jennings}},\
  and\ \bibinfo {author} {\bibfnamefont {T.}~\bibnamefont {Rudolph}},\ }\href
  {https://dx.doi.org/10.1088/1367-2630/18/2/023037} {\bibfield  {journal}
  {\bibinfo  {journal} {NJP}\ }\textbf {\bibinfo {volume} {18}},\ \bibinfo
  {pages} {023037} (\bibinfo {year} {2016})},\ \Eprint
  {https://arxiv.org/abs/1508.02720} {arXiv:1508.02720} \BibitemShut {NoStop}%
\bibitem [{\citenamefont {Woods}\ and\ \citenamefont
  {Horodecki}(2023)}]{woods_autonomous_2023}%
  \BibitemOpen
  \bibfield  {author} {\bibinfo {author} {\bibfnamefont {M.~P.}\ \bibnamefont
  {Woods}}\ and\ \bibinfo {author} {\bibfnamefont {M.}~\bibnamefont
  {Horodecki}},\ }\href {https://link.aps.org/doi/10.1103/PhysRevX.13.011016}
  {\bibfield  {journal} {\bibinfo  {journal} {Phys. Rev. X}\ }\textbf {\bibinfo
  {volume} {13}},\ \bibinfo {pages} {011016} (\bibinfo {year} {2023})},\
  \Eprint {https://arxiv.org/abs/1912.05562} {arXiv:1912.05562} \BibitemShut
  {NoStop}%
\bibitem [{\citenamefont {Ball}\ \emph {et~al.}(2016)\citenamefont {Ball},
  \citenamefont {Oliver},\ and\ \citenamefont {Biercuk}}]{ball_role_2016}%
  \BibitemOpen
  \bibfield  {author} {\bibinfo {author} {\bibfnamefont {H.}~\bibnamefont
  {Ball}}, \bibinfo {author} {\bibfnamefont {W.~D.}\ \bibnamefont {Oliver}},\
  and\ \bibinfo {author} {\bibfnamefont {M.~J.}\ \bibnamefont {Biercuk}},\
  }\href {https://doi.org/10.1038/npjqi.2016.33} {\bibfield  {journal}
  {\bibinfo  {journal} {npj Quantum Inf.}\ }\textbf {\bibinfo {volume} {2}},\
  \bibinfo {pages} {1} (\bibinfo {year} {2016})},\ \Eprint
  {https://arxiv.org/abs/1602.04551} {arXiv:1602.04551} \BibitemShut {NoStop}%
\bibitem [{\citenamefont {Xuereb}\ \emph {et~al.}(2023)\citenamefont {Xuereb},
  \citenamefont {Erker}, \citenamefont {Meier}, \citenamefont {Mitchison},\
  and\ \citenamefont {Huber}}]{xuereb_impact_2023}%
  \BibitemOpen
  \bibfield  {author} {\bibinfo {author} {\bibfnamefont {J.}~\bibnamefont
  {Xuereb}}, \bibinfo {author} {\bibfnamefont {P.}~\bibnamefont {Erker}},
  \bibinfo {author} {\bibfnamefont {F.}~\bibnamefont {Meier}}, \bibinfo
  {author} {\bibfnamefont {M.~T.}\ \bibnamefont {Mitchison}},\ and\ \bibinfo
  {author} {\bibfnamefont {M.}~\bibnamefont {Huber}},\ }\href
  {https://doi.org/10.1103/PhysRevLett.131.160204} {\bibfield  {journal}
  {\bibinfo  {journal} {Phys. Rev. Lett.}\ }\textbf {\bibinfo {volume} {131}},\
  \bibinfo {pages} {160204} (\bibinfo {year} {2023})}\BibitemShut {NoStop}%
\bibitem [{\citenamefont {Hauge}\ and\ \citenamefont
  {St{\o}vneng}(1989)}]{hauge_tunneling_1989}%
  \BibitemOpen
  \bibfield  {author} {\bibinfo {author} {\bibfnamefont {E.~H.}\ \bibnamefont
  {Hauge}}\ and\ \bibinfo {author} {\bibfnamefont {J.~A.}\ \bibnamefont
  {St{\o}vneng}},\ }\href {https://link.aps.org/doi/10.1103/RevModPhys.61.917}
  {\bibfield  {journal} {\bibinfo  {journal} {Reviews of Modern Phys.}\
  }\textbf {\bibinfo {volume} {61}},\ \bibinfo {pages} {917} (\bibinfo {year}
  {1989})}\BibitemShut {NoStop}%
\bibitem [{\citenamefont {Silva}\ \emph {et~al.}(2023)\citenamefont {Silva},
  \citenamefont {Nurgalieva},\ and\ \citenamefont
  {Wilming}}]{silva_ticking_2023}%
  \BibitemOpen
  \bibfield  {author} {\bibinfo {author} {\bibfnamefont {R.}~\bibnamefont
  {Silva}}, \bibinfo {author} {\bibfnamefont {N.}~\bibnamefont {Nurgalieva}},\
  and\ \bibinfo {author} {\bibfnamefont {H.}~\bibnamefont {Wilming}},\ }\href
  {http://arxiv.org/abs/2306.01829} {\bibinfo {title} {Ticking clocks in
  quantum theory}} (\bibinfo {year} {2023}),\ \bibinfo {note}
  {arXiv:2306.01829}\BibitemShut {NoStop}%
\bibitem [{\citenamefont {Pietzonka}(2022)}]{pietzonka_classical_2022}%
  \BibitemOpen
  \bibfield  {author} {\bibinfo {author} {\bibfnamefont {P.}~\bibnamefont
  {Pietzonka}},\ }\href
  {https://link.aps.org/doi/10.1103/PhysRevLett.128.130606} {\bibfield
  {journal} {\bibinfo  {journal} {Phys. Rev. Letters}\ }\textbf {\bibinfo
  {volume} {128}},\ \bibinfo {pages} {130606} (\bibinfo {year} {2022})},\
  \Eprint {https://arxiv.org/abs/2110.02213} {arXiv:2110.02213} \BibitemShut
  {NoStop}%
\bibitem [{\citenamefont {Wen}\ \emph {et~al.}(2020)\citenamefont {Wen},
  \citenamefont {Ares}, \citenamefont {Schupp}, \citenamefont {Pei},
  \citenamefont {Briggs},\ and\ \citenamefont {Laird}}]{Wen2020}%
  \BibitemOpen
  \bibfield  {author} {\bibinfo {author} {\bibfnamefont {Y.}~\bibnamefont
  {Wen}}, \bibinfo {author} {\bibfnamefont {N.}~\bibnamefont {Ares}}, \bibinfo
  {author} {\bibfnamefont {F.~J.}\ \bibnamefont {Schupp}}, \bibinfo {author}
  {\bibfnamefont {T.}~\bibnamefont {Pei}}, \bibinfo {author} {\bibfnamefont
  {G.~A.~D.}\ \bibnamefont {Briggs}},\ and\ \bibinfo {author} {\bibfnamefont
  {E.~A.}\ \bibnamefont {Laird}},\ }\href
  {https://doi.org/10.1038/s41567-019-0683-5} {\bibfield  {journal} {\bibinfo
  {journal} {Nature Physics}\ }\textbf {\bibinfo {volume} {16}},\ \bibinfo
  {pages} {75} (\bibinfo {year} {2020})},\ \Eprint
  {https://arxiv.org/abs/1903.04474} {arXiv:1903.04474} \BibitemShut {NoStop}%
\bibitem [{\citenamefont {Urgell}\ \emph {et~al.}(2020)\citenamefont {Urgell},
  \citenamefont {Yang}, \citenamefont {{De Bonis}}, \citenamefont {Samanta},
  \citenamefont {Esplandiu}, \citenamefont {Dong}, \citenamefont {Jin},\ and\
  \citenamefont {Bachtold}}]{Urgell2020}%
  \BibitemOpen
  \bibfield  {author} {\bibinfo {author} {\bibfnamefont {C.}~\bibnamefont
  {Urgell}}, \bibinfo {author} {\bibfnamefont {W.}~\bibnamefont {Yang}},
  \bibinfo {author} {\bibfnamefont {S.~L.}\ \bibnamefont {{De Bonis}}},
  \bibinfo {author} {\bibfnamefont {C.}~\bibnamefont {Samanta}}, \bibinfo
  {author} {\bibfnamefont {M.~J.}\ \bibnamefont {Esplandiu}}, \bibinfo {author}
  {\bibfnamefont {Q.}~\bibnamefont {Dong}}, \bibinfo {author} {\bibfnamefont
  {Y.}~\bibnamefont {Jin}},\ and\ \bibinfo {author} {\bibfnamefont
  {A.}~\bibnamefont {Bachtold}},\ }\href
  {https://doi.org/10.1038/s41567-019-0682-6} {\bibfield  {journal} {\bibinfo
  {journal} {Nat. Phys.}\ }\textbf {\bibinfo {volume} {16}},\ \bibinfo {pages}
  {32} (\bibinfo {year} {2020})},\ \Eprint {https://arxiv.org/abs/1903.04892}
  {arXiv:1903.04892} \BibitemShut {NoStop}%
\bibitem [{\citenamefont {Jenkins}(2013)}]{jenkins_self-oscillation_2013}%
  \BibitemOpen
  \bibfield  {author} {\bibinfo {author} {\bibfnamefont {A.}~\bibnamefont
  {Jenkins}},\ }\href
  {https://www.sciencedirect.com/science/article/pii/S0370157312004073}
  {\bibfield  {journal} {\bibinfo  {journal} {Phys. Reports}\ }\bibinfo
  {series} {Self-oscillation},\ \textbf {\bibinfo {volume} {525}},\ \bibinfo
  {pages} {167} (\bibinfo {year} {2013})},\ \Eprint
  {https://arxiv.org/abs/1109.6640} {arXiv:1109.6640} \BibitemShut {NoStop}%
\bibitem [{\citenamefont {Bachtold}\ \emph {et~al.}(2022)\citenamefont
  {Bachtold}, \citenamefont {Moser},\ and\ \citenamefont
  {Dykman}}]{Bachtold2022}%
  \BibitemOpen
  \bibfield  {author} {\bibinfo {author} {\bibfnamefont {A.}~\bibnamefont
  {Bachtold}}, \bibinfo {author} {\bibfnamefont {J.}~\bibnamefont {Moser}},\
  and\ \bibinfo {author} {\bibfnamefont {M.~I.}\ \bibnamefont {Dykman}},\
  }\href {https://doi.org/10.1103/RevModPhys.94.045005} {\bibfield  {journal}
  {\bibinfo  {journal} {Rev. Mod. Phys.}\ }\textbf {\bibinfo {volume} {94}},\
  \bibinfo {pages} {045005} (\bibinfo {year} {2022})},\ \Eprint
  {https://arxiv.org/abs/2202.01819} {arXiv:2202.01819} \BibitemShut {NoStop}%
\bibitem [{\citenamefont {Gorelik}\ \emph {et~al.}(1998)\citenamefont
  {Gorelik}, \citenamefont {Isacsson}, \citenamefont {Voinova}, \citenamefont
  {Kasemo}, \citenamefont {Shekhter},\ and\ \citenamefont
  {Jonson}}]{Gorelik1998}%
  \BibitemOpen
  \bibfield  {author} {\bibinfo {author} {\bibfnamefont {L.~Y.}\ \bibnamefont
  {Gorelik}}, \bibinfo {author} {\bibfnamefont {A.}~\bibnamefont {Isacsson}},
  \bibinfo {author} {\bibfnamefont {M.~V.}\ \bibnamefont {Voinova}}, \bibinfo
  {author} {\bibfnamefont {B.}~\bibnamefont {Kasemo}}, \bibinfo {author}
  {\bibfnamefont {R.~I.}\ \bibnamefont {Shekhter}},\ and\ \bibinfo {author}
  {\bibfnamefont {M.}~\bibnamefont {Jonson}},\ }\href
  {https://doi.org/10.1103/physrevlett.80.4526} {\bibfield  {journal} {\bibinfo
   {journal} {Physical Review Letters}\ }\textbf {\bibinfo {volume} {80}},\
  \bibinfo {pages} {4526–4529} (\bibinfo {year} {1998})},\ \Eprint
  {https://arxiv.org/abs/9711196} {arXiv:9711196} \BibitemShut {NoStop}%
\bibitem [{\citenamefont {Park}\ \emph {et~al.}(2000)\citenamefont {Park},
  \citenamefont {Park}, \citenamefont {Lim}, \citenamefont {Anderson},
  \citenamefont {Alivisatos},\ and\ \citenamefont {McEuen}}]{Park_2000}%
  \BibitemOpen
  \bibfield  {author} {\bibinfo {author} {\bibfnamefont {H.}~\bibnamefont
  {Park}}, \bibinfo {author} {\bibfnamefont {J.}~\bibnamefont {Park}}, \bibinfo
  {author} {\bibfnamefont {A.~K.~L.}\ \bibnamefont {Lim}}, \bibinfo {author}
  {\bibfnamefont {E.~H.}\ \bibnamefont {Anderson}}, \bibinfo {author}
  {\bibfnamefont {A.~P.}\ \bibnamefont {Alivisatos}},\ and\ \bibinfo {author}
  {\bibfnamefont {P.~L.}\ \bibnamefont {McEuen}},\ }\href
  {https://doi.org/10.1038/35024031} {\bibfield  {journal} {\bibinfo  {journal}
  {Nature}\ }\textbf {\bibinfo {volume} {407}},\ \bibinfo {pages} {57}
  (\bibinfo {year} {2000})}\BibitemShut {NoStop}%
\bibitem [{\citenamefont {Erbe}\ \emph {et~al.}(2001)\citenamefont {Erbe},
  \citenamefont {Weiss}, \citenamefont {Zwerger},\ and\ \citenamefont
  {Blick}}]{erbe_nanomechanical_2001}%
  \BibitemOpen
  \bibfield  {author} {\bibinfo {author} {\bibfnamefont {A.}~\bibnamefont
  {Erbe}}, \bibinfo {author} {\bibfnamefont {C.}~\bibnamefont {Weiss}},
  \bibinfo {author} {\bibfnamefont {W.}~\bibnamefont {Zwerger}},\ and\ \bibinfo
  {author} {\bibfnamefont {R.~H.}\ \bibnamefont {Blick}},\ }\href
  {https://link.aps.org/doi/10.1103/PhysRevLett.87.096106} {\bibfield
  {journal} {\bibinfo  {journal} {Phys. Rev. Letters}\ }\textbf {\bibinfo
  {volume} {87}},\ \bibinfo {pages} {096106} (\bibinfo {year} {2001})},\
  \Eprint {https://arxiv.org/abs/0011429} {arXiv:0011429} \BibitemShut
  {NoStop}%
\bibitem [{\citenamefont {Pistolesi}\ and\ \citenamefont
  {Fazio}(2006)}]{Pistolesi2006}%
  \BibitemOpen
  \bibfield  {author} {\bibinfo {author} {\bibfnamefont {F.}~\bibnamefont
  {Pistolesi}}\ and\ \bibinfo {author} {\bibfnamefont {R.}~\bibnamefont
  {Fazio}},\ }\href {https://doi.org/10.1088/1367-2630/8/7/113} {\bibfield
  {journal} {\bibinfo  {journal} {New Journal of Physics}\ }\textbf {\bibinfo
  {volume} {8}},\ \bibinfo {pages} {113} (\bibinfo {year} {2006})},\ \Eprint
  {https://arxiv.org/abs/0608538} {arXiv:0608538} \BibitemShut {NoStop}%
\bibitem [{\citenamefont {Novotn\'y}\ \emph {et~al.}(2003)\citenamefont
  {Novotn\'y}, \citenamefont {Donarini},\ and\ \citenamefont
  {Jauho}}]{Novotn2003}%
  \BibitemOpen
  \bibfield  {author} {\bibinfo {author} {\bibfnamefont {T.~c.~v.}\
  \bibnamefont {Novotn\'y}}, \bibinfo {author} {\bibfnamefont {A.}~\bibnamefont
  {Donarini}},\ and\ \bibinfo {author} {\bibfnamefont {A.-P.}\ \bibnamefont
  {Jauho}},\ }\href {https://doi.org/10.1103/PhysRevLett.90.256801} {\bibfield
  {journal} {\bibinfo  {journal} {Phys. Rev. Lett.}\ }\textbf {\bibinfo
  {volume} {90}},\ \bibinfo {pages} {256801} (\bibinfo {year} {2003})},\
  \Eprint {https://arxiv.org/abs/0301441} {arXiv:0301441} \BibitemShut
  {NoStop}%
\bibitem [{\citenamefont {Novotn\'y}\ \emph {et~al.}(2004)\citenamefont
  {Novotn\'y}, \citenamefont {Donarini}, \citenamefont {Flindt},\ and\
  \citenamefont {Jauho}}]{Novotn2004}%
  \BibitemOpen
  \bibfield  {author} {\bibinfo {author} {\bibfnamefont {T.~c.~v.}\
  \bibnamefont {Novotn\'y}}, \bibinfo {author} {\bibfnamefont {A.}~\bibnamefont
  {Donarini}}, \bibinfo {author} {\bibfnamefont {C.}~\bibnamefont {Flindt}},\
  and\ \bibinfo {author} {\bibfnamefont {A.-P.}\ \bibnamefont {Jauho}},\ }\href
  {https://doi.org/10.1103/PhysRevLett.92.248302} {\bibfield  {journal}
  {\bibinfo  {journal} {Phys. Rev. Lett.}\ }\textbf {\bibinfo {volume} {92}},\
  \bibinfo {pages} {248302} (\bibinfo {year} {2004})},\ \Eprint
  {https://arxiv.org/abs/0402597} {arXiv:0402597} \BibitemShut {NoStop}%
\bibitem [{\citenamefont {Blanter}\ \emph {et~al.}(2004)\citenamefont
  {Blanter}, \citenamefont {Usmani},\ and\ \citenamefont
  {Nazarov}}]{Blanter2004}%
  \BibitemOpen
  \bibfield  {author} {\bibinfo {author} {\bibfnamefont {Y.~M.}\ \bibnamefont
  {Blanter}}, \bibinfo {author} {\bibfnamefont {O.}~\bibnamefont {Usmani}},\
  and\ \bibinfo {author} {\bibfnamefont {Y.~V.}\ \bibnamefont {Nazarov}},\
  }\href {https://doi.org/10.1103/PhysRevLett.93.136802} {\bibfield  {journal}
  {\bibinfo  {journal} {Phys. Rev. Lett.}\ }\textbf {\bibinfo {volume} {93}},\
  \bibinfo {pages} {136802} (\bibinfo {year} {2004})},\ \Eprint
  {https://arxiv.org/abs/0404615} {arXiv:0404615} \BibitemShut {NoStop}%
\bibitem [{\citenamefont {Blanter}\ \emph {et~al.}(2005)\citenamefont
  {Blanter}, \citenamefont {Usmani},\ and\ \citenamefont
  {Nazarov}}]{Blanter2004err}%
  \BibitemOpen
  \bibfield  {author} {\bibinfo {author} {\bibfnamefont {Y.~M.}\ \bibnamefont
  {Blanter}}, \bibinfo {author} {\bibfnamefont {O.}~\bibnamefont {Usmani}},\
  and\ \bibinfo {author} {\bibfnamefont {Y.~V.}\ \bibnamefont {Nazarov}},\
  }\href {https://doi.org/10.1103/PhysRevLett.94.049904} {\bibfield  {journal}
  {\bibinfo  {journal} {Phys. Rev. Lett.}\ }\textbf {\bibinfo {volume} {94}},\
  \bibinfo {pages} {049904} (\bibinfo {year} {2005})}\BibitemShut {NoStop}%
\bibitem [{\citenamefont {Clerk}\ and\ \citenamefont
  {Bennett}(2005)}]{Clerk2005}%
  \BibitemOpen
  \bibfield  {author} {\bibinfo {author} {\bibfnamefont {A.~A.}\ \bibnamefont
  {Clerk}}\ and\ \bibinfo {author} {\bibfnamefont {S.}~\bibnamefont
  {Bennett}},\ }\href {https://doi.org/10.1088/1367-2630/7/1/238} {\bibfield
  {journal} {\bibinfo  {journal} {New Journal of Physics}\ }\textbf {\bibinfo
  {volume} {7}},\ \bibinfo {pages} {238} (\bibinfo {year} {2005})},\ \Eprint
  {https://arxiv.org/abs/0507646} {arXiv:0507646} \BibitemShut {NoStop}%
\bibitem [{\citenamefont {Bennett}\ and\ \citenamefont
  {Clerk}(2006)}]{Bennett2006}%
  \BibitemOpen
  \bibfield  {author} {\bibinfo {author} {\bibfnamefont {S.~D.}\ \bibnamefont
  {Bennett}}\ and\ \bibinfo {author} {\bibfnamefont {A.~A.}\ \bibnamefont
  {Clerk}},\ }\href {https://doi.org/10.1103/PhysRevB.74.201301} {\bibfield
  {journal} {\bibinfo  {journal} {Phys. Rev. B}\ }\textbf {\bibinfo {volume}
  {74}},\ \bibinfo {pages} {201301} (\bibinfo {year} {2006})}\BibitemShut
  {NoStop}%
\bibitem [{\citenamefont {Usmani}\ \emph {et~al.}(2007)\citenamefont {Usmani},
  \citenamefont {Blanter},\ and\ \citenamefont {Nazarov}}]{Usmani2007}%
  \BibitemOpen
  \bibfield  {author} {\bibinfo {author} {\bibfnamefont {O.}~\bibnamefont
  {Usmani}}, \bibinfo {author} {\bibfnamefont {Y.~M.}\ \bibnamefont
  {Blanter}},\ and\ \bibinfo {author} {\bibfnamefont {Y.~V.}\ \bibnamefont
  {Nazarov}},\ }\href {https://doi.org/10.1103/PhysRevB.75.195312} {\bibfield
  {journal} {\bibinfo  {journal} {Phys. Rev. B}\ }\textbf {\bibinfo {volume}
  {75}},\ \bibinfo {pages} {195312} (\bibinfo {year} {2007})},\ \Eprint
  {https://arxiv.org/abs/0603017} {arXiv:0603017} \BibitemShut {NoStop}%
\bibitem [{\citenamefont {W\"{a}chtler}\ \emph {et~al.}(2019)\citenamefont
  {W\"{a}chtler}, \citenamefont {Strasberg}, \citenamefont {Klapp},
  \citenamefont {Schaller},\ and\ \citenamefont {Jarzynski}}]{Wachtler_2019}%
  \BibitemOpen
  \bibfield  {author} {\bibinfo {author} {\bibfnamefont {C.~W.}\ \bibnamefont
  {W\"{a}chtler}}, \bibinfo {author} {\bibfnamefont {P.}~\bibnamefont
  {Strasberg}}, \bibinfo {author} {\bibfnamefont {S.~H.~L.}\ \bibnamefont
  {Klapp}}, \bibinfo {author} {\bibfnamefont {G.}~\bibnamefont {Schaller}},\
  and\ \bibinfo {author} {\bibfnamefont {C.}~\bibnamefont {Jarzynski}},\ }\href
  {https://doi.org/10.1088/1367-2630/ab2727} {\bibfield  {journal} {\bibinfo
  {journal} {New Journal of Physics}\ }\textbf {\bibinfo {volume} {21}},\
  \bibinfo {pages} {073009} (\bibinfo {year} {2019})},\ \Eprint
  {https://arxiv.org/abs/1902.08174} {arXiv:1902.08174} \BibitemShut {NoStop}%
\bibitem [{\citenamefont {Strasberg}\ \emph {et~al.}(2021)\citenamefont
  {Strasberg}, \citenamefont {W{\"{a}}chtler},\ and\ \citenamefont
  {Schaller}}]{Strasberg2021}%
  \BibitemOpen
  \bibfield  {author} {\bibinfo {author} {\bibfnamefont {P.}~\bibnamefont
  {Strasberg}}, \bibinfo {author} {\bibfnamefont {C.~W.}\ \bibnamefont
  {W{\"{a}}chtler}},\ and\ \bibinfo {author} {\bibfnamefont {G.}~\bibnamefont
  {Schaller}},\ }\href {https://doi.org/10.1103/PhysRevLett.126.180605}
  {\bibfield  {journal} {\bibinfo  {journal} {Phys. Rev. Lett.}\ }\textbf
  {\bibinfo {volume} {126}},\ \bibinfo {pages} {180605} (\bibinfo {year}
  {2021})},\ \Eprint {https://arxiv.org/abs/2101.05027} {arXiv:2101.05027}
  \BibitemShut {NoStop}%
\bibitem [{\citenamefont {Culhane}\ \emph {et~al.}(2022)\citenamefont
  {Culhane}, \citenamefont {Mitchison},\ and\ \citenamefont
  {Goold}}]{Culhane_2022}%
  \BibitemOpen
  \bibfield  {author} {\bibinfo {author} {\bibfnamefont {O.}~\bibnamefont
  {Culhane}}, \bibinfo {author} {\bibfnamefont {M.~T.}\ \bibnamefont
  {Mitchison}},\ and\ \bibinfo {author} {\bibfnamefont {J.}~\bibnamefont
  {Goold}},\ }\href {https://doi.org/10.1103/PhysRevE.106.L032104} {\bibfield
  {journal} {\bibinfo  {journal} {Phys. Rev. E}\ }\textbf {\bibinfo {volume}
  {106}},\ \bibinfo {pages} {L032104} (\bibinfo {year} {2022})},\ \Eprint
  {https://arxiv.org/abs/2201.07819} {arXiv:2201.07819} \BibitemShut {NoStop}%
\bibitem [{\citenamefont {Pearson}\ \emph {et~al.}(2021)\citenamefont
  {Pearson}, \citenamefont {Guryanova}, \citenamefont {Erker}, \citenamefont
  {Laird}, \citenamefont {Briggs}, \citenamefont {Huber},\ and\ \citenamefont
  {Ares}}]{Pearson2021}%
  \BibitemOpen
  \bibfield  {author} {\bibinfo {author} {\bibfnamefont {A.~N.}\ \bibnamefont
  {Pearson}}, \bibinfo {author} {\bibfnamefont {Y.}~\bibnamefont {Guryanova}},
  \bibinfo {author} {\bibfnamefont {P.}~\bibnamefont {Erker}}, \bibinfo
  {author} {\bibfnamefont {E.~A.}\ \bibnamefont {Laird}}, \bibinfo {author}
  {\bibfnamefont {G.~A.~D.}\ \bibnamefont {Briggs}}, \bibinfo {author}
  {\bibfnamefont {M.}~\bibnamefont {Huber}},\ and\ \bibinfo {author}
  {\bibfnamefont {N.}~\bibnamefont {Ares}},\ }\href
  {https://doi.org/10.1103/PhysRevX.11.021029} {\bibfield  {journal} {\bibinfo
  {journal} {Phys. Rev. X}\ }\textbf {\bibinfo {volume} {11}},\ \bibinfo
  {pages} {021029} (\bibinfo {year} {2021})},\ \Eprint
  {https://arxiv.org/abs/2006.08670} {arXiv:2006.08670} \BibitemShut {NoStop}%
\bibitem [{\citenamefont {He}\ \emph {et~al.}(2023)\citenamefont {He},
  \citenamefont {Pakkiam}, \citenamefont {Gangat}, \citenamefont {Kewming},
  \citenamefont {Milburn},\ and\ \citenamefont {Fedorov}}]{he_quantum_2023}%
  \BibitemOpen
  \bibfield  {author} {\bibinfo {author} {\bibfnamefont {X.}~\bibnamefont
  {He}}, \bibinfo {author} {\bibfnamefont {P.}~\bibnamefont {Pakkiam}},
  \bibinfo {author} {\bibfnamefont {A.~A.}\ \bibnamefont {Gangat}}, \bibinfo
  {author} {\bibfnamefont {M.~J.}\ \bibnamefont {Kewming}}, \bibinfo {author}
  {\bibfnamefont {G.~J.}\ \bibnamefont {Milburn}},\ and\ \bibinfo {author}
  {\bibfnamefont {A.}~\bibnamefont {Fedorov}},\ }\href
  {https://doi.org/10.1103/PhysRevApplied.20.034038} {\bibfield  {journal}
  {\bibinfo  {journal} {Phys. Rev. Appl.}\ }\textbf {\bibinfo {volume} {20}},\
  \bibinfo {pages} {034038} (\bibinfo {year} {2023})}\BibitemShut {NoStop}%
\bibitem [{\citenamefont {Allan}(1966)}]{Allan_1966}%
  \BibitemOpen
  \bibfield  {author} {\bibinfo {author} {\bibfnamefont {D.}~\bibnamefont
  {Allan}},\ }\href {https://doi.org/10.1109/proc.1966.4634} {\bibfield
  {journal} {\bibinfo  {journal} {Proceedings of the {IEEE}}\ }\textbf
  {\bibinfo {volume} {54}},\ \bibinfo {pages} {221} (\bibinfo {year}
  {1966})}\BibitemShut {NoStop}%
\bibitem [{\citenamefont {Gardiner}\ and\ \citenamefont
  {Zoller}()}]{GardinerZoller}%
  \BibitemOpen
  \bibfield  {author} {\bibinfo {author} {\bibfnamefont {C.~W.}\ \bibnamefont
  {Gardiner}}\ and\ \bibinfo {author} {\bibfnamefont {P.}~\bibnamefont
  {Zoller}},\ }\href@noop {} {\emph {\bibinfo {title} {{Quantum Noise}}}},\
  \bibinfo {edition} {3rd}\ ed.\ (\bibinfo  {publisher} {Springer},\ \bibinfo
  {address} {Berlin})\BibitemShut {NoStop}%
\bibitem [{\citenamefont {Gardiner}(2009)}]{Gardiner}%
  \BibitemOpen
  \bibfield  {author} {\bibinfo {author} {\bibfnamefont {C.}~\bibnamefont
  {Gardiner}},\ }\href
  {https://www.ebook.de/de/product/7853242/crispin_gardiner_stochastic_methods.html}
  {\emph {\bibinfo {title} {Stochastic Methods}}}\ (\bibinfo  {publisher}
  {Springer Berlin Heidelberg},\ \bibinfo {year} {2009})\BibitemShut {NoStop}%
\bibitem [{\citenamefont {Vigneau}\ \emph {et~al.}(2023)\citenamefont
  {Vigneau}, \citenamefont {Fedele}, \citenamefont {Chatterjee}, \citenamefont
  {Reilly}, \citenamefont {Kuemmeth}, \citenamefont {Gonzalez-Zalba},
  \citenamefont {Laird},\ and\ \citenamefont {Ares}}]{vigneau_probing_2023}%
  \BibitemOpen
  \bibfield  {author} {\bibinfo {author} {\bibfnamefont {F.}~\bibnamefont
  {Vigneau}}, \bibinfo {author} {\bibfnamefont {F.}~\bibnamefont {Fedele}},
  \bibinfo {author} {\bibfnamefont {A.}~\bibnamefont {Chatterjee}}, \bibinfo
  {author} {\bibfnamefont {D.}~\bibnamefont {Reilly}}, \bibinfo {author}
  {\bibfnamefont {F.}~\bibnamefont {Kuemmeth}}, \bibinfo {author}
  {\bibfnamefont {M.~F.}\ \bibnamefont {Gonzalez-Zalba}}, \bibinfo {author}
  {\bibfnamefont {E.}~\bibnamefont {Laird}},\ and\ \bibinfo {author}
  {\bibfnamefont {N.}~\bibnamefont {Ares}},\ }\href
  {https://doi.org/10.1063/5.0088229} {\bibfield  {journal} {\bibinfo
  {journal} {Applied Phys. Reviews}\ }\textbf {\bibinfo {volume} {10}},\
  \bibinfo {pages} {021305} (\bibinfo {year} {2023})}\BibitemShut {NoStop}%
\bibitem [{\citenamefont {Clerk}\ \emph {et~al.}(2010)\citenamefont {Clerk},
  \citenamefont {Devoret}, \citenamefont {Girvin}, \citenamefont {Marquardt},\
  and\ \citenamefont {Schoelkopf}}]{clerk_introduction_2010}%
  \BibitemOpen
  \bibfield  {author} {\bibinfo {author} {\bibfnamefont {A.~A.}\ \bibnamefont
  {Clerk}}, \bibinfo {author} {\bibfnamefont {M.~H.}\ \bibnamefont {Devoret}},
  \bibinfo {author} {\bibfnamefont {S.~M.}\ \bibnamefont {Girvin}}, \bibinfo
  {author} {\bibfnamefont {F.}~\bibnamefont {Marquardt}},\ and\ \bibinfo
  {author} {\bibfnamefont {R.~J.}\ \bibnamefont {Schoelkopf}},\ }\href
  {https://link.aps.org/doi/10.1103/RevModPhys.82.1155} {\bibfield  {journal}
  {\bibinfo  {journal} {Reviews of Modern Phys.}\ }\textbf {\bibinfo {volume}
  {82}},\ \bibinfo {pages} {1155} (\bibinfo {year} {2010})},\ \Eprint
  {https://arxiv.org/abs/0810.4729} {arXiv:0810.4729} \BibitemShut {NoStop}%
\bibitem [{\citenamefont {Ryndyk}(2016)}]{ryndyk2016}%
  \BibitemOpen
  \bibfield  {author} {\bibinfo {author} {\bibfnamefont {D.~A.}\ \bibnamefont
  {Ryndyk}},\ }\bibinfo {title} {Green functions},\ in\ \href@noop {} {\emph
  {\bibinfo {booktitle} {Theory of Quantum transport at nanoscale an
  introduction}}}\ (\bibinfo  {publisher} {Springer International Publishing},\
  \bibinfo {year} {2016})\BibitemShut {NoStop}%
\bibitem [{\citenamefont {Gurvitz}\ and\ \citenamefont
  {Prager}(1996)}]{Gurvitz1996}%
  \BibitemOpen
  \bibfield  {author} {\bibinfo {author} {\bibfnamefont {S.~A.}\ \bibnamefont
  {Gurvitz}}\ and\ \bibinfo {author} {\bibfnamefont {Y.~S.}\ \bibnamefont
  {Prager}},\ }\href {https://doi.org/10.1103/PhysRevB.53.15932} {\bibfield
  {journal} {\bibinfo  {journal} {Phys. Rev. B}\ }\textbf {\bibinfo {volume}
  {53}},\ \bibinfo {pages} {15932} (\bibinfo {year} {1996})}\BibitemShut
  {NoStop}%
\bibitem [{\citenamefont {Risken}(1996)}]{risken_fokker-planck_1996}%
  \BibitemOpen
  \bibfield  {author} {\bibinfo {author} {\bibfnamefont {H.}~\bibnamefont
  {Risken}},\ }\href {https://link.springer.com/10.1007/978-3-642-61544-3}
  {{\selectlanguage {English}\emph {\bibinfo {title} {The {Fokker}-{Planck}
  {Equation}: {Methods} of {Solution} and {Applications}}}}},\ edited by\
  \bibinfo {editor} {\bibfnamefont {H.}~\bibnamefont {Haken}},\ \bibinfo
  {series} {Springer {Series} in {Synergetics}}, Vol.~\bibinfo {volume} {18}\
  (\bibinfo  {publisher} {Springer},\ \bibinfo {address} {Berlin, Heidelberg},\
  \bibinfo {year} {1996})\BibitemShut {NoStop}%
\bibitem [{\citenamefont {Tabanera-Bravo}\ \emph {et~al.}(2022)\citenamefont
  {Tabanera-Bravo}, \citenamefont {Vigneau}, \citenamefont {Monsel},
  \citenamefont {Aggarwal}, \citenamefont {Bresque}, \citenamefont {Fedele},
  \citenamefont {Cerisola}, \citenamefont {Briggs}, \citenamefont {Anders},
  \citenamefont {Auf{\`e}ves}, \citenamefont {Parrondo},\ and\ \citenamefont
  {Ares}}]{tabanera-bravo_stability_2022}%
  \BibitemOpen
  \bibfield  {author} {\bibinfo {author} {\bibfnamefont {J.}~\bibnamefont
  {Tabanera-Bravo}}, \bibinfo {author} {\bibfnamefont {F.}~\bibnamefont
  {Vigneau}}, \bibinfo {author} {\bibfnamefont {J.}~\bibnamefont {Monsel}},
  \bibinfo {author} {\bibfnamefont {K.}~\bibnamefont {Aggarwal}}, \bibinfo
  {author} {\bibfnamefont {L.}~\bibnamefont {Bresque}}, \bibinfo {author}
  {\bibfnamefont {F.}~\bibnamefont {Fedele}}, \bibinfo {author} {\bibfnamefont
  {F.}~\bibnamefont {Cerisola}}, \bibinfo {author} {\bibfnamefont {G.~A.~D.}\
  \bibnamefont {Briggs}}, \bibinfo {author} {\bibfnamefont {J.}~\bibnamefont
  {Anders}}, \bibinfo {author} {\bibfnamefont {A.}~\bibnamefont {Auf{\`e}ves}},
  \bibinfo {author} {\bibfnamefont {J.~M.~R.}\ \bibnamefont {Parrondo}},\ and\
  \bibinfo {author} {\bibfnamefont {N.}~\bibnamefont {Ares}},\ }\href
  {http://arxiv.org/abs/2211.04074} {\bibinfo {title} {Stability of
  long-sustained oscillations induced by electron tunneling}} (\bibinfo {year}
  {2022}),\ \bibinfo {note} {arXiv:2211.04074}\BibitemShut {NoStop}%
\bibitem [{\citenamefont {Kewming}\ \emph {et~al.}(2022)\citenamefont
  {Kewming}, \citenamefont {Mitchison},\ and\ \citenamefont
  {Landi}}]{kewming_diverging_2022}%
  \BibitemOpen
  \bibfield  {author} {\bibinfo {author} {\bibfnamefont {M.~J.}\ \bibnamefont
  {Kewming}}, \bibinfo {author} {\bibfnamefont {M.~T.}\ \bibnamefont
  {Mitchison}},\ and\ \bibinfo {author} {\bibfnamefont {G.~T.}\ \bibnamefont
  {Landi}},\ }\href {https://link.aps.org/doi/10.1103/PhysRevA.106.033707}
  {\bibfield  {journal} {\bibinfo  {journal} {Phys. Rev. A}\ }\textbf {\bibinfo
  {volume} {106}},\ \bibinfo {pages} {033707} (\bibinfo {year} {2022})},\
  \Eprint {https://arxiv.org/abs/2205.02622} {arXiv:2205.02622} \BibitemShut
  {NoStop}%
\bibitem [{\citenamefont {Aminzare}\ \emph {et~al.}(2019)\citenamefont
  {Aminzare}, \citenamefont {Holmes},\ and\ \citenamefont
  {Srivastava}}]{Aminzare2019}%
  \BibitemOpen
  \bibfield  {author} {\bibinfo {author} {\bibfnamefont {Z.}~\bibnamefont
  {Aminzare}}, \bibinfo {author} {\bibfnamefont {P.}~\bibnamefont {Holmes}},\
  and\ \bibinfo {author} {\bibfnamefont {V.}~\bibnamefont {Srivastava}},\ }in\
  \href {https://doi.org/10.1109/cdc40024.2019.9030112} {\emph {\bibinfo
  {booktitle} {2019 {IEEE} 58th Conference on Decision and Control ({CDC})}}}\
  (\bibinfo  {publisher} {{IEEE}},\ \bibinfo {year} {2019})\BibitemShut
  {NoStop}%
\bibitem [{\citenamefont {Cox}(1967)}]{Cox_1967}%
  \BibitemOpen
  \bibfield  {author} {\bibinfo {author} {\bibfnamefont {D.~R.}\ \bibnamefont
  {Cox}},\ }\href@noop {} {\emph {\bibinfo {title} {Renewal theory}}}\
  (\bibinfo  {publisher} {Methuen},\ \bibinfo {year} {1967})\BibitemShut
  {NoStop}%
\bibitem [{\citenamefont {Radaelli}\ \emph {et~al.}(2023)\citenamefont
  {Radaelli}, \citenamefont {Landi}, \citenamefont {Modi},\ and\ \citenamefont
  {Binder}}]{radaelli_fisher_2023}%
  \BibitemOpen
  \bibfield  {author} {\bibinfo {author} {\bibfnamefont {M.}~\bibnamefont
  {Radaelli}}, \bibinfo {author} {\bibfnamefont {G.~T.}\ \bibnamefont {Landi}},
  \bibinfo {author} {\bibfnamefont {K.}~\bibnamefont {Modi}},\ and\ \bibinfo
  {author} {\bibfnamefont {F.~C.}\ \bibnamefont {Binder}},\ }\href
  {http://arxiv.org/abs/2206.00463} {\bibfield  {journal} {\bibinfo  {journal}
  {NJP}\ }\textbf {\bibinfo {volume} {25}},\ \bibinfo {pages} {053037}
  (\bibinfo {year} {2023})},\ \Eprint {https://arxiv.org/abs/2206.00463}
  {arXiv:2206.00463} \BibitemShut {NoStop}%
\bibitem [{\citenamefont {Altaie}\ \emph {et~al.}(2022)\citenamefont {Altaie},
  \citenamefont {Hodgson},\ and\ \citenamefont {Beige}}]{Altaie2022}%
  \BibitemOpen
  \bibfield  {author} {\bibinfo {author} {\bibfnamefont {M.~B.}\ \bibnamefont
  {Altaie}}, \bibinfo {author} {\bibfnamefont {D.}~\bibnamefont {Hodgson}},\
  and\ \bibinfo {author} {\bibfnamefont {A.}~\bibnamefont {Beige}},\ }\href
  {https://doi.org/10.3389/fphy.2022.897305} {\bibfield  {journal} {\bibinfo
  {journal} {Frontiers in Physics}\ }\textbf {\bibinfo {volume} {10}},\
  \bibinfo {pages} {897305} (\bibinfo {year} {2022})}\BibitemShut {NoStop}%
\bibitem [{\citenamefont {Brenes}\ \emph {et~al.}(2020)\citenamefont {Brenes},
  \citenamefont {Mendoza-Arenas}, \citenamefont {Purkayastha}, \citenamefont
  {Mitchison}, \citenamefont {Clark},\ and\ \citenamefont
  {Goold}}]{Brenes_2020}%
  \BibitemOpen
  \bibfield  {author} {\bibinfo {author} {\bibfnamefont {M.}~\bibnamefont
  {Brenes}}, \bibinfo {author} {\bibfnamefont {J.~J.}\ \bibnamefont
  {Mendoza-Arenas}}, \bibinfo {author} {\bibfnamefont {A.}~\bibnamefont
  {Purkayastha}}, \bibinfo {author} {\bibfnamefont {M.~T.}\ \bibnamefont
  {Mitchison}}, \bibinfo {author} {\bibfnamefont {S.~R.}\ \bibnamefont
  {Clark}},\ and\ \bibinfo {author} {\bibfnamefont {J.}~\bibnamefont {Goold}},\
  }\href {https://doi.org/10.1103/PhysRevX.10.031040} {\bibfield  {journal}
  {\bibinfo  {journal} {Phys. Rev. X}\ }\textbf {\bibinfo {volume} {10}},\
  \bibinfo {pages} {031040} (\bibinfo {year} {2020})},\ \Eprint
  {https://arxiv.org/abs/1912.02053} {arXiv:1912.02053} \BibitemShut {NoStop}%
\bibitem [{\citenamefont {Lacerda}\ \emph {et~al.}(2023)\citenamefont
  {Lacerda}, \citenamefont {Purkayastha}, \citenamefont {Kewming},
  \citenamefont {Landi},\ and\ \citenamefont {Goold}}]{lacerda_quantum_2023}%
  \BibitemOpen
  \bibfield  {author} {\bibinfo {author} {\bibfnamefont {A.~M.}\ \bibnamefont
  {Lacerda}}, \bibinfo {author} {\bibfnamefont {A.}~\bibnamefont
  {Purkayastha}}, \bibinfo {author} {\bibfnamefont {M.}~\bibnamefont
  {Kewming}}, \bibinfo {author} {\bibfnamefont {G.~T.}\ \bibnamefont {Landi}},\
  and\ \bibinfo {author} {\bibfnamefont {J.}~\bibnamefont {Goold}},\ }\href
  {https://link.aps.org/doi/10.1103/PhysRevB.107.195117} {\bibfield  {journal}
  {\bibinfo  {journal} {Phys. Rev. B}\ }\textbf {\bibinfo {volume} {107}},\
  \bibinfo {pages} {195117} (\bibinfo {year} {2023})},\ \Eprint
  {https://arxiv.org/abs/2206.01090} {arXiv:2206.01090} \BibitemShut {NoStop}%
\bibitem [{\citenamefont {Van~Vu}\ and\ \citenamefont
  {Hasegawa}(2019)}]{Van_Vu_2019}%
  \BibitemOpen
  \bibfield  {author} {\bibinfo {author} {\bibfnamefont {T.}~\bibnamefont
  {Van~Vu}}\ and\ \bibinfo {author} {\bibfnamefont {Y.}~\bibnamefont
  {Hasegawa}},\ }\href {https://doi.org/10.1103/PhysRevE.100.032130} {\bibfield
   {journal} {\bibinfo  {journal} {Phys. Rev. E}\ }\textbf {\bibinfo {volume}
  {100}},\ \bibinfo {pages} {032130} (\bibinfo {year} {2019})},\ \Eprint
  {https://arxiv.org/abs/1901.05715} {arXiv:1901.05715} \BibitemShut {NoStop}%
\bibitem [{\citenamefont {Chan}\ \emph {et~al.}(2015)\citenamefont {Chan},
  \citenamefont {Lee},\ and\ \citenamefont
  {Gopalakrishnan}}]{chan_limit-cycle_2015}%
  \BibitemOpen
  \bibfield  {author} {\bibinfo {author} {\bibfnamefont {C.-K.}\ \bibnamefont
  {Chan}}, \bibinfo {author} {\bibfnamefont {T.~E.}\ \bibnamefont {Lee}},\ and\
  \bibinfo {author} {\bibfnamefont {S.}~\bibnamefont {Gopalakrishnan}},\ }\href
  {https://link.aps.org/doi/10.1103/PhysRevA.91.051601} {\bibfield  {journal}
  {\bibinfo  {journal} {Phys. Rev. A}\ }\textbf {\bibinfo {volume} {91}},\
  \bibinfo {pages} {051601} (\bibinfo {year} {2015})},\ \Eprint
  {https://arxiv.org/abs/1501.00979} {arXiv:1501.00979} \BibitemShut {NoStop}%
\bibitem [{\citenamefont {Iemini}\ \emph {et~al.}(2018)\citenamefont {Iemini},
  \citenamefont {Russomanno}, \citenamefont {Keeling}, \citenamefont
  {Schir{\`o}}, \citenamefont {Dalmonte},\ and\ \citenamefont
  {Fazio}}]{iemini_boundary_2018}%
  \BibitemOpen
  \bibfield  {author} {\bibinfo {author} {\bibfnamefont {F.}~\bibnamefont
  {Iemini}}, \bibinfo {author} {\bibfnamefont {A.}~\bibnamefont {Russomanno}},
  \bibinfo {author} {\bibfnamefont {J.}~\bibnamefont {Keeling}}, \bibinfo
  {author} {\bibfnamefont {M.}~\bibnamefont {Schir{\`o}}}, \bibinfo {author}
  {\bibfnamefont {M.}~\bibnamefont {Dalmonte}},\ and\ \bibinfo {author}
  {\bibfnamefont {R.}~\bibnamefont {Fazio}},\ }\href
  {https://link.aps.org/doi/10.1103/PhysRevLett.121.035301} {\bibfield
  {journal} {\bibinfo  {journal} {Phys. Rev. Letters}\ }\textbf {\bibinfo
  {volume} {121}},\ \bibinfo {pages} {035301} (\bibinfo {year} {2018})},\
  \Eprint {https://arxiv.org/abs/1708.05014} {arXiv:1708.05014} \BibitemShut
  {NoStop}%
\bibitem [{\citenamefont {Bu{\v c}a}\ \emph {et~al.}(2019)\citenamefont {Bu{\v
  c}a}, \citenamefont {Tindall},\ and\ \citenamefont
  {Jaksch}}]{buca_non-stationary_2019}%
  \BibitemOpen
  \bibfield  {author} {\bibinfo {author} {\bibfnamefont {B.}~\bibnamefont
  {Bu{\v c}a}}, \bibinfo {author} {\bibfnamefont {J.}~\bibnamefont {Tindall}},\
  and\ \bibinfo {author} {\bibfnamefont {D.}~\bibnamefont {Jaksch}},\ }\href
  {https://www.nature.com/articles/s41467-019-09757-y} {\bibfield  {journal}
  {\bibinfo  {journal} {Nature Communications}\ }\textbf {\bibinfo {volume}
  {10}},\ \bibinfo {pages} {1730} (\bibinfo {year} {2019})},\ \Eprint
  {https://arxiv.org/abs/1804.06744} {arXiv:1804.06744} \BibitemShut {NoStop}%
\bibitem [{\citenamefont {Guarnieri}\ \emph {et~al.}(2022)\citenamefont
  {Guarnieri}, \citenamefont {Mitchison}, \citenamefont {Purkayastha},
  \citenamefont {Jaksch}, \citenamefont {Bu{\v c}a},\ and\ \citenamefont
  {Goold}}]{guarnieri_time_2022}%
  \BibitemOpen
  \bibfield  {author} {\bibinfo {author} {\bibfnamefont {G.}~\bibnamefont
  {Guarnieri}}, \bibinfo {author} {\bibfnamefont {M.~T.}\ \bibnamefont
  {Mitchison}}, \bibinfo {author} {\bibfnamefont {A.}~\bibnamefont
  {Purkayastha}}, \bibinfo {author} {\bibfnamefont {D.}~\bibnamefont {Jaksch}},
  \bibinfo {author} {\bibfnamefont {B.}~\bibnamefont {Bu{\v c}a}},\ and\
  \bibinfo {author} {\bibfnamefont {J.}~\bibnamefont {Goold}},\ }\href
  {https://link.aps.org/doi/10.1103/PhysRevA.106.022209} {\bibfield  {journal}
  {\bibinfo  {journal} {Phys. Rev. A}\ }\textbf {\bibinfo {volume} {106}},\
  \bibinfo {pages} {022209} (\bibinfo {year} {2022})},\ \Eprint
  {https://arxiv.org/abs/2104.13402} {arXiv:2104.13402} \BibitemShut {NoStop}%
\bibitem [{\citenamefont {Wilczek}(2012)}]{wilczek_quantum_2012}%
  \BibitemOpen
  \bibfield  {author} {\bibinfo {author} {\bibfnamefont {F.}~\bibnamefont
  {Wilczek}},\ }\href {https://link.aps.org/doi/10.1103/PhysRevLett.109.160401}
  {\bibfield  {journal} {\bibinfo  {journal} {Phys. Rev. Letters}\ }\textbf
  {\bibinfo {volume} {109}},\ \bibinfo {pages} {160401} (\bibinfo {year}
  {2012})},\ \Eprint {https://arxiv.org/abs/1202.2539} {arXiv:1202.2539}
  \BibitemShut {NoStop}%
\bibitem [{\citenamefont {Sacha}\ and\ \citenamefont
  {Zakrzewski}(2018)}]{sacha_time_2018}%
  \BibitemOpen
  \bibfield  {author} {\bibinfo {author} {\bibfnamefont {K.}~\bibnamefont
  {Sacha}}\ and\ \bibinfo {author} {\bibfnamefont {J.}~\bibnamefont
  {Zakrzewski}},\ }\href {http://arxiv.org/abs/1704.03735} {\bibfield
  {journal} {\bibinfo  {journal} {Reports on Progress in Phys.}\ }\textbf
  {\bibinfo {volume} {81}},\ \bibinfo {pages} {016401} (\bibinfo {year}
  {2018})},\ \Eprint {https://arxiv.org/abs/1704.03735} {arXiv:1704.03735}
  \BibitemShut {NoStop}%
\bibitem [{\citenamefont {Zaletel}\ \emph {et~al.}(2023)\citenamefont
  {Zaletel}, \citenamefont {Lukin}, \citenamefont {Monroe}, \citenamefont
  {Nayak}, \citenamefont {Wilczek},\ and\ \citenamefont
  {Yao}}]{zaletel_colloquium_2023}%
  \BibitemOpen
  \bibfield  {author} {\bibinfo {author} {\bibfnamefont {M.~P.}\ \bibnamefont
  {Zaletel}}, \bibinfo {author} {\bibfnamefont {M.}~\bibnamefont {Lukin}},
  \bibinfo {author} {\bibfnamefont {C.}~\bibnamefont {Monroe}}, \bibinfo
  {author} {\bibfnamefont {C.}~\bibnamefont {Nayak}}, \bibinfo {author}
  {\bibfnamefont {F.}~\bibnamefont {Wilczek}},\ and\ \bibinfo {author}
  {\bibfnamefont {N.~Y.}\ \bibnamefont {Yao}},\ }\href
  {https://doi.org/10.1103/RevModPhys.95.031001} {\bibfield  {journal}
  {\bibinfo  {journal} {Rev. Mod. Phys.}\ }\textbf {\bibinfo {volume} {95}},\
  \bibinfo {pages} {031001} (\bibinfo {year} {2023})}\BibitemShut {NoStop}%
\bibitem [{\citenamefont {Landi}\ \emph {et~al.}(2023)\citenamefont {Landi},
  \citenamefont {Kewming}, \citenamefont {Mitchison},\ and\ \citenamefont
  {Potts}}]{landi_current_2023}%
  \BibitemOpen
  \bibfield  {author} {\bibinfo {author} {\bibfnamefont {G.~T.}\ \bibnamefont
  {Landi}}, \bibinfo {author} {\bibfnamefont {M.~J.}\ \bibnamefont {Kewming}},
  \bibinfo {author} {\bibfnamefont {M.~T.}\ \bibnamefont {Mitchison}},\ and\
  \bibinfo {author} {\bibfnamefont {P.~P.}\ \bibnamefont {Potts}},\ }\href
  {http://arxiv.org/abs/2303.04270} {\bibinfo {title} {Current fluctuations in
  open quantum systems: {Bridging} the gap between quantum continuous
  measurements and full counting statistics}} (\bibinfo {year} {2023}),\
  \bibinfo {note} {arXiv:2303.04270}\BibitemShut {NoStop}%
\end{thebibliography}%

\end{document}